\pdfoutput=1

\newcommand{\degr}{$^\circ$}
\newcommand{\caproman}[1]{\uppercase\expandafter{\romannumeral#1}}

\newcommand{\BGmail}{bgottsch\,@\,fas.harvard.edu}

\documentclass{article}

\title{\bf On the nuclear halo of a proton pencil beam\\stopping in water}
\author{Bernard Gottschalk\thanks{Harvard University Laboratory for Particle Physics and Cosmology, 18 Hammond St., Cambridge, MA 02138, USA (corresponding author, \BGmail)}, ~Ethan W. Cascio\thanks{Francis H. Burr Proton Therapy Center, Mass. General Hospital,
30 Fruit Street, Boston, MA, USA 02114}, ~Juliane Daartz\thanks{Francis H. Burr Proton Therapy Center} ~and Miles S. Wagner\thanks{Mevion Medical Systems Inc., 300 Foster St.
Littleton, MA 01460, USA}}

\usepackage{amsmath}
\usepackage{latexsym}
\usepackage{graphicx}

\begin{document}

\maketitle

\begin{abstract}
\noindent
The dose distribution of a proton beam stopping in water has components due to basic physics and may have others from beam contamination. We propose the concise terms {\em core} for the primary beam, {\em halo} (cf. Pedroni et al. \cite{pedroniPencil}) for the low dose region from charged secondaries, {\em aura} for  the low dose region from neutrals, and {\em spray} for beam contamination. 

We have measured the dose distribution in a water tank at 177\,MeV under conditions where spray, therefore radial asymmetry, is negligible. We used an ADCL calibrated thimble chamber and a Faraday cup calibrated integral beam monitor so as to obtain immediately the absolute dose per proton. We took depth scans at fixed distances from the beam centroid rather than radial scans at fixed depths. That minimizes the signal range for each scan and better reveals the structure of the core and halo.

Transitions from core to halo to aura are already discernible in the raw data. The halo has components attributable to coherent and incoherent nuclear reactions. Due to elastic and inelastic scattering by the nuclear force, the Bragg peak persists to radii larger than can be accounted for by Moli\`ere single scattering. The radius of the incoherent component, a dose bump around midrange, agrees with the kinematics of knockout reactions.

We have fitted the data in two ways. The first is algebraic or model dependent (MD) as far as possible, and has 25 parameters. The second, using 2D cubic spline regression, is model independent (MI). Optimal parameterization for treatment planning will probably be a hybrid of the two, and will of course require measurements at several incident energies.

The MD fit to the core term resembles that of the PSI group \cite{pedroniPencil}, which has been widely emulated. However, we replace their $T(w)$, a mass stopping power which mixes electromagnetic (EM) and nuclear effects, with one that is purely EM, arguing that protons that do {\em not} undergo hard single scatters continue to lose energy according to the Beth-Bloch formula. If that is correct, it is no longer necessary to measure $T(w)$, and the dominant role played by the `Bragg peak chamber' (BPW) vanishes.

For mathematical and other details we will refer to \cite{Gottschalk2014}, a long technical report of this project.
\end{abstract}

%\clearpage\tableofcontents

\clearpage
\section{Introduction}
This paper concerns a topic which has gained interest in the last decade owing to the increased use of pencil beam scanning (PBS). The Paul Scherrer Institute (PSI) group coined the term {\em halo} for the cloud of nuclear secon\-daries that surrounds a pencil beam stopping in matter, and proposed a model whose parameters they found by experiment \cite{pedroniPencil}. To put their model in perspective we quote (emphasis ours):
\begin{quote}
`These are first preliminary estimates of the 'first moment' of the spatial lateral dose distribution due to nuclear interaction products and large angle Coulomb scattering of a proton pencil beam in water. Unfortunately the dosimetric method (the smallness of the effect) is
not sufficiently precise for extracting information on the higher moments (the shape of the beam halo). Monte Carlo studies could help us further on this subject. {\em The point we want to stress is that this issue matters if one wants to predict absolute dose directly in the dose calculation.}'
\end{quote}
Since then some twenty papers on the topic have appeared, each addressing specific clinical issues such as commissioning a treatment planning system (TPS) for a particular treatment beam. When a model is called for, all use variants of the PSI model. However, even though each paper covers some aspect of the halo, one is ultimately left without a comprehensive picture.

Our own interest began with the `Bragg Peak chamber' (BPC) problem. Central to the PSI model is the radial integral of the dose distribution measured with a large circular plane-parallel ion chamber, a so-called BPC. The PSI group used a 4\,cm radius BPC, later found to be too small. Because of the ubiquity of the PSI model, BPC-related considerations, e.g. correcting  for missing dose experimentally \cite{Anand2012} or by Monte-Carlo simulation \cite{Clasie2012} account for a considerable fraction of the literature. 

Using medium energy nuclear physics, in particular energy and momentum conservation, we were able to show (Sec.\,\ref{sec:physics}) that the most energetic charged secondaries reach a radius approximately 1/3 the range of the incident beam. Therefore, commercial BPCs are indeed too small at the higher PBS energies. However, thinking about that problem led us to ask why, in describing a cylindrically symmetric dose distribution, the {\em radial integral} of that distribution should play such a large role. That led us to question the PSI model itself and the usefulness of BPC measurements (Sec.\,\ref{sec:discussion}).

We therefore undertook measurement and parameterization of the halo without using a BPC. Because of restrictions in time, money and manpower, we used existing equipment with minor modifications, in a flexible and already well understood test beam. We followed Sawakuchi et al. \cite{Sawakuchi2010}, using a single beam in a water tank with a small cylindrical ionization chamber (IC), but with two important changes:
\begin{itemize}
\item Scans in depth at selected radii rather than radial scans at selected depths. That reduces the signal range for most scans, and better reveals the structure of the halo. 
\item In-house electronics chain rather than the built-in water tank electronics to permit using a plane-parallel ion chamber (PPIC) integral beam monitor instead of a reference IC.
\end{itemize}
The result was, for the first time, a comprehensive absolute measurement of the halo, albeit at just one energy (Sec.\,\ref{sec:measurement}).

We subsequently fit the data (Table\,\ref{tbl:dmlg}) in a model-dependent (MD) as well as a model-independent (MI) fashion (Sec.\,\ref{sec:fits}). These are preliminary exercises to explore what can be done. We regard parameterizing the full dose distribution efficiently for clinical purposes as a still unsolved problem, beyond the scope of this paper. It will probably involve both MD and MI components. Of course, measurements or Monte Carlo (MC) simulations at a number of energies will be needed.

\begin{figure}[p]
\centering\includegraphics[width=4.97in,height=3.5in]{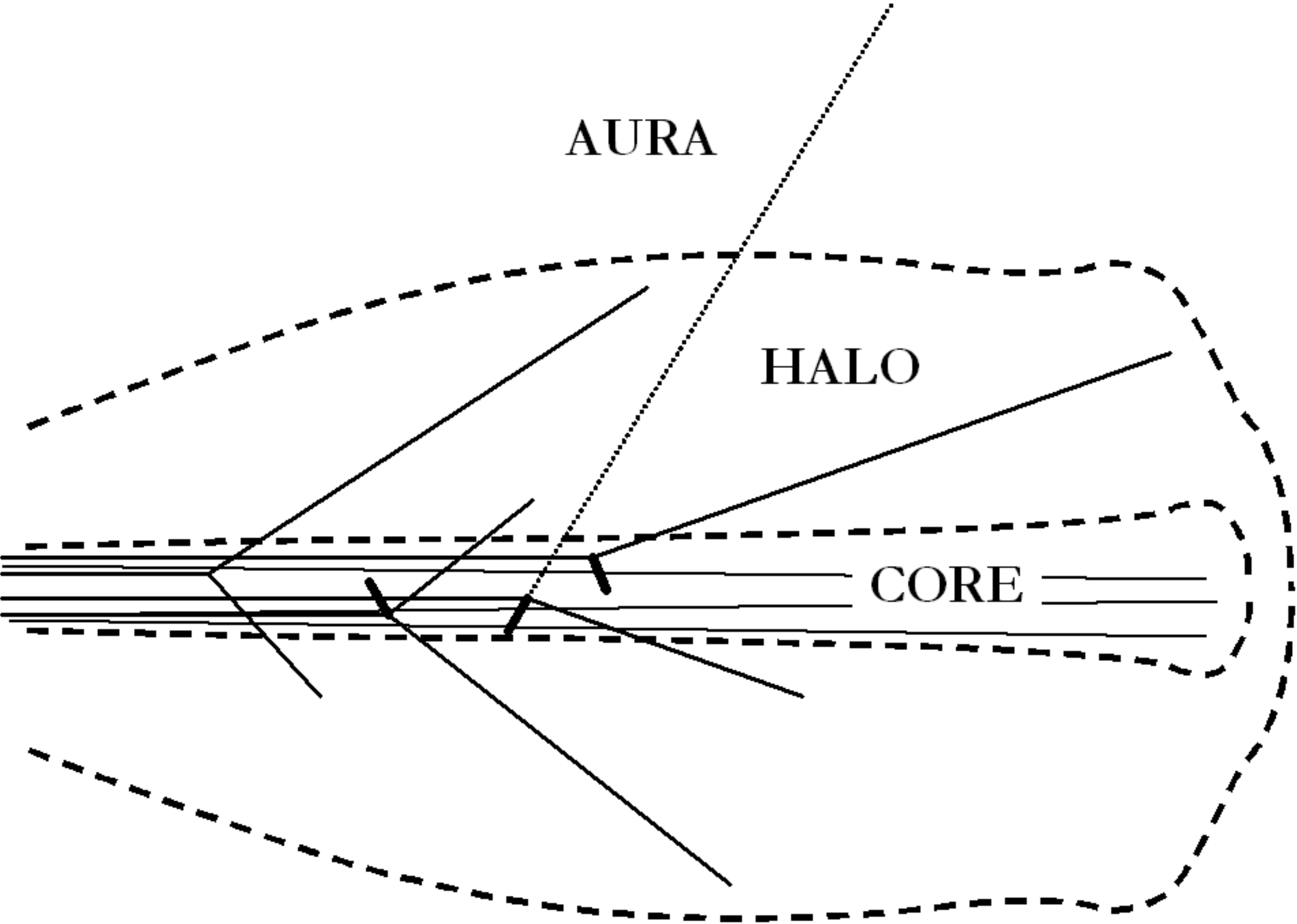}% 1.42
\caption{Core, halo and aura with schematic reactions (from left) $^1$H(p,p)p, $^{16}$O(p,2p)$^{15}$N, $^{16}$O(p,pn)$^{15}$O and $^{16}$O(p,p)$^{16}$O. Recoil nuclei ranges are exaggerated. The dashed lines are 10\% and 0.01\% isodoses drawn to scale.\label{fig:haloReactions}}
\end{figure}

\begin{figure}[p]
\centering\includegraphics[width=5.60in,height=3.5in]{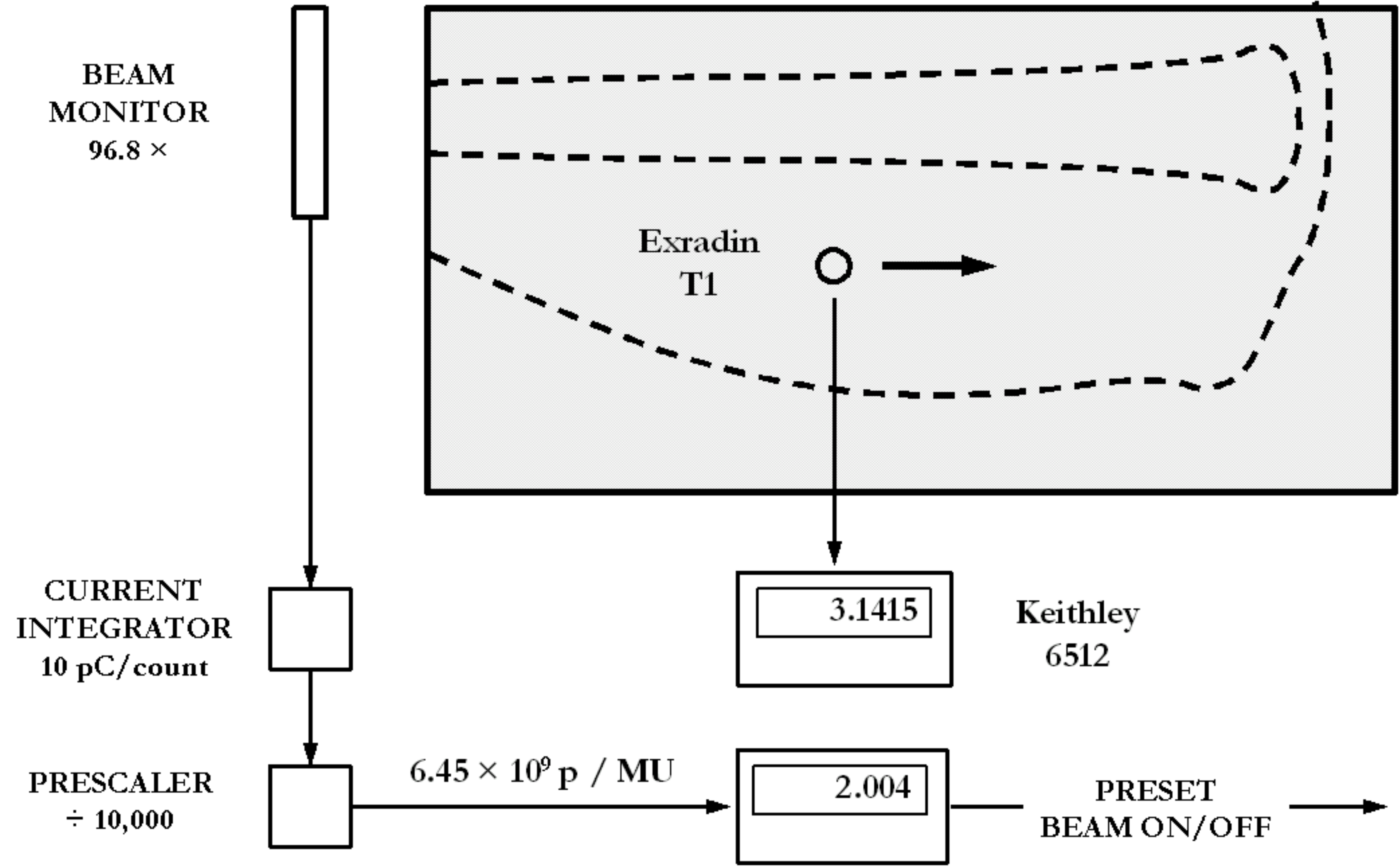}% 1.62
\caption{Experimental setup with water tank in top view. The T1 dosimeter stem is vertical (into the plane of the figure).\label{fig:haloSetup}}
\end{figure}

\begin{figure}[p]
\centering\includegraphics[width=4.35in,height=3.5in]{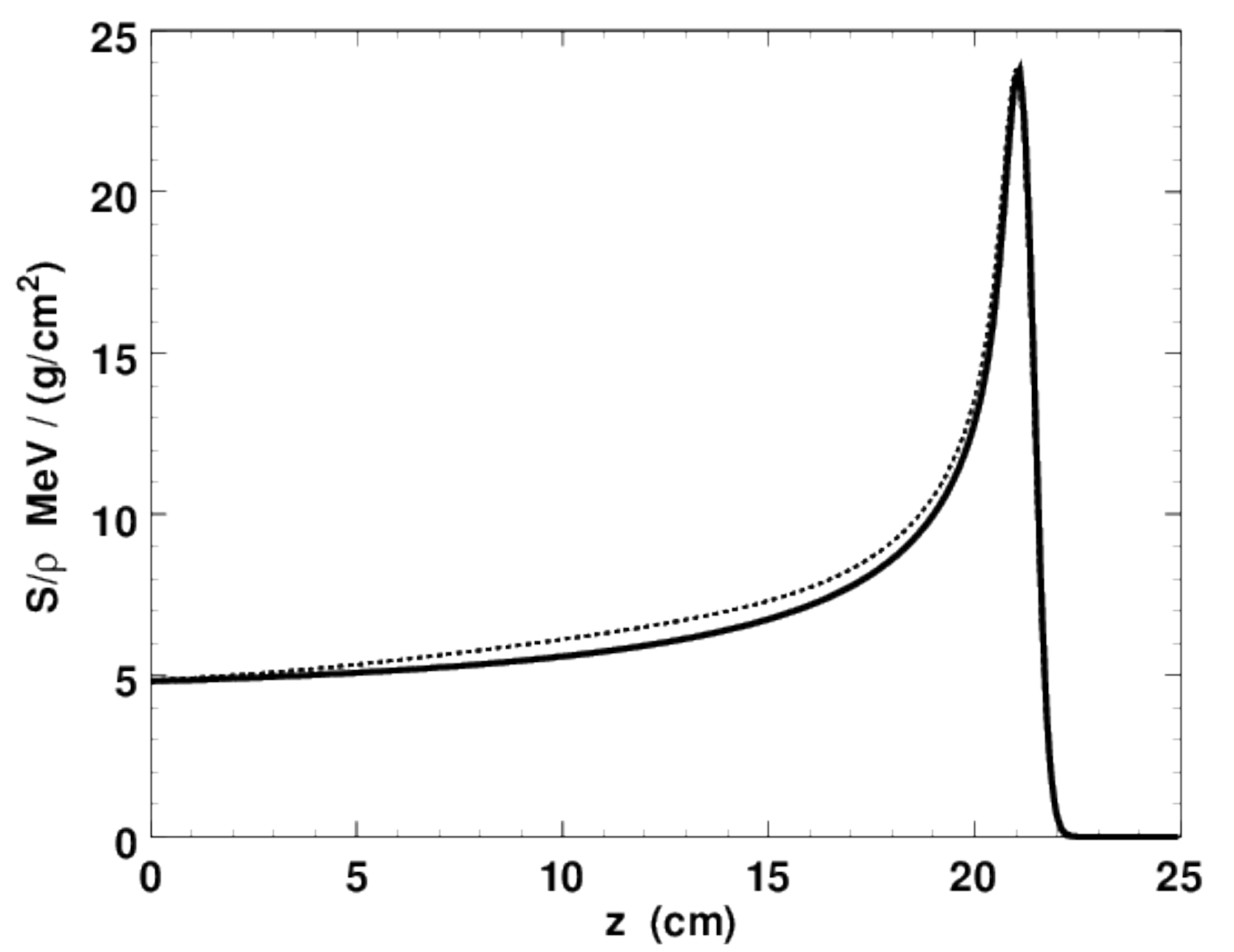}% 1.242
\caption{$S_\mathrm{mixed}/\rho$ from the MD fit (dashed line) and $(1-cz)\times S_\mathrm{em}/\rho$ with $c$ adjusted so the two peaks are equal (solid line). The difference at midrange is 10\%.\label{fig:S_EM}}
\end{figure}

This paper seeks to convey a qualitative understanding of the halo. To minimize mathematics we rely heavily on \cite{Gottschalk2014}, a very long technical report of the same project. Relevant nuclear reactions, the MD and MI fits, and a review of the literature are treated there at far greater length. On the other hand Figs.\,\ref{fig:haloReactions}, \ref{fig:haloSetup}, \ref{fig:depthScans}, \ref{fig:S_EM} and accompanying text are new here, as are (thanks to a referee) several citations. We hope this broad-brush treatment will be useful to the student, and will give even the expert food for thought.

Finally, though both were done at Massachusetts General Hospital's Francis H. Burr Proton Therapy Center (hereinafter `Burr Center'), there is no overlap in purpose, method, beam line or authorship between this and the earlier work of Clasie et al. \cite{Clasie2012}, which used the PSI model and provided data for the Astroid TPS. That was an independent study.

\section{Anatomy of the Dose Distribution}
To standardize the terminology we propose the terms {\em core}, {\em halo}, {\em aura} and {\em spray} for components of the dose distribution. Fig.\,\ref{fig:haloReactions} illustrates the first three. They overlap, but each has distinct spatial characteristics. These proposed categories will be justified by what follows.

\subsection{Core}
When protons enter a water tank they undergo myriad interactions with atomic electrons, which slow them down, and with atomic nuclei, which scatter them through small angles. Most protons stop solely as a result of these multiple electromagnetic (EM) interactions. We call them, and only them, {\em primaries} and they make up the core or `primary beam' which is what we normally think of as `the pencil beam'.\,\footnote{~Our definition of primaries (and by implication, secondaries) is not universal. The Monte Carlo MCNPX, for instance, calls any elastically scattered proton a primary. That makes one of the final protons from elastic scattering on free hydrogen a primary and the other a secondary, even though they have identical energy/angle characteristics.}

Taken together, the Bethe-Bloch theory of slowing down \cite{janni82,icru49} and the Moli\`ere theory of multiple Coulomb scattering (MCS) \cite{moliere1,moliere2,bethe,mcsbg} lead to Fermi-Eyges (FE) theory \cite{transport2012}, a well tested transport theory of the core. By `transport' we mean that, given seven incident `phase space parameters' for instance the beam kinetic energy $T$ and its rms size, rms angular spread and emittance in each transverse direction $x,y$ we can find those same parameters at any depth $z$. The incident beam must be approximately Gaussian in $x,y$ and $\theta_x,\theta_y$ (projected angles to the beam direction).\footnote{~In water the MCS angle is always small. Thus $\theta_x,\theta_y$ can also be regarded as slopes.} If various materials are traversed in turn, each must be homogeneous in $x$ and $y$ (`mixed slab' geometry). Under those conditions, all distributions characterizing the core at any $z$ will also be Gaussian.

Near the stopping depth FE theory breaks down because of energy straggling, which blurs the relation between depth and proton energy. We need an additional parameter, a range spread, to find the complete primary dose distribution. That parameter combines the irreducible range straggling \cite{janni82} with the site specific beam energy spread (expressed in terms of range) in quadrature, and determines the width of the Bragg peak \cite{bortfeld}. 

In summary, given eight parameters of the incident beam, we can find corresponding parameters of the core at any depth by a deterministic calculation, even if the beam passes through homogeneous slabs on the way to the water. We have deliberately skipped the fairly complicated math, which can be found elsewhere \cite{pedroniPencil,Gottschalk2014,transport2012}.

The key fact omitted so far is that {\em primaries are lost} due to single `hard' scatters on H or O. Four such scatters are shown schematically in Fig.\,\ref{fig:haloReactions}. They may involve either the Coulomb or the nuclear force, and may be elastic, inelastic or nonelastic (Sec.\,\ref{sec:physics}). Therefore we must multiply the primary fluence by a `primary attrition function' which depends on $z$. For example, at 177\,MeV roughly 20\% of the incident protons suffer a {\em nonelastic} nuclear reaction before stopping \cite{janni82}, and that represents about 12\% of the beam energy \cite{Gottschalk2014}. Single {\em elastic} and {\em inelastic} scatters also contribute to the halo, and to the attrition function.

\subsection{Halo}
The halo consists of the {\em charged secondaries} from hard single scatters. These are EM elastic scatters from H and O (the `single scattering tail' of the Moli\`ere distribution), nuclear elastic scatters from H and O, nuclear inelastic scatters from O, and nonelastic scatters from O. We will show that the halo radius is roughly one-third the beam range. Therefore, depending on target size, the halo may deliver dose where it is either wanted or unwanted. 

Being rather diffuse, the halo hardly affects the {\em shape} of the high-dose region. However, as we have stated, it contains a significant fraction of the integrated dose (energy). If we start with a central pencil and then surround it with neighboring pencils, the central dose will increase due to the overlapping halos. As the neighbors get further out, the increase will eventually level off. This behavior, implied by our quote from \cite{pedroniPencil}, is often characterized by a {\em field size factor} (FSF).

The halo in turn will be seen to have two components, which probably depend differently on beam energy:

\subsubsection{Bragg peak}
Fig.\,5 of \cite{pedroniPencil} shows a noticeable peak even from frame D. The scattering angle implied by the geometry of that frame is far greater than can be accounted for by the Moli\`ere single scattering tail. What is going on here is coherent scattering of primaries by the {\em nuclear} force (Sec.\,\ref{sec:coherent}, Fig.\,\ref{fig:Gerstein3}).

\subsubsection{Midrange bump}
Several studies e.g. \cite{Lin2014} suggest, and our depth-dose measurements (Fig.\,\ref{fig:dmlin}) confirm, a broad dose enhancement at large radii just beyond midrange of the incident beam. That is also supported by kinematic arguments (Sec.\,\ref{sec:size}), and is due to large-angle protons from  elastic scattering on H and nonelastic scattering on O.

\subsection{Aura}
The aura consists of neutral secondaries ($\gamma-$rays and neutrons) from inelastic and nonelastic nuclear interactions, and all the particles they set in motion. It is very large, pervading the patient, treatment room, shielding and facility and depositing unwanted dose in all. Its relative contribution to the high-dose region is negligible. Its behavior at large distances depends strongly on the materials traversed e.g. shielding. However, we will see it only in the water immediately outside the core and halo, whose geometry is well defined.

\subsection{Spray}
We propose this term for unwanted beam that comes from upstream and can in principle be avoided. Together, the spray and halo make up what is often called the `low-dose region'. (The halo itself, in our definition of that term, is due to unavoidable nuclear physics.) Spray is tolerable if it can be parameterized and/or its contribution to the low-dose region is sufficiently small. However, it certainly complicates commissioning a TPS for pencil beam scanning e.g. \cite{Zhu2013,LinTOPAS2014}. Three different sources appear in the literature to date:

\subsubsection{Degraders near the patient} 
PSI \cite{pedroniPencil} uses uniform degraders near the patient as range shifters. They are encompassed by FE theory as regards the core, and simply broaden it. Howeve, some but not all secondaries from the degraders also reach the patient. PSI \cite{pedroniPencil} proposes a simple empirical model to account for them. They can be avoided by using a variable energy machine or by degrading and re-analyzing the beam far from the patient.

\subsubsection{Beam profile monitors} 
The M.D. Anderson group \cite{Sawakuchi2010} uses tungsten wire chambers rather far from the patient as beam profile monitors (BPMs). Protons passing through the tungsten scatter strongly, especially at low energies, and can be treated approximately as a `secondary core' \cite{Zhu2013} which, at low energies, actually exceeds the nuclear halo. BPM spray can be avoided by using very thin and uniform segmented ion chambers, which have recently been commercialized (Pyramid Technical Consultants, 1050 Waltham St, Lexington, MA 02421).

\subsubsection{Beam pipe walls}
The UPenn group \cite{Lin2014,Lin2013novel,Lin2013} reports spray due (they speculate) to protons scattered from the beam pipe lining their focusing magnets. Those protons (we speculate) are caused by scattering in the first BPM. If the beamline design requires focusing magnets downstream of a vacuum window or BPM, this type of spray may be hard to mitigate because magnet cost increases rapidly, and scan speed decreases, with aperture. Also, this spray may depend on the transverse position of the beam. It may therefore vary during a scan and be more difficult to parameterize. 

\section{Nuclear Physics of the Halo}\label{sec:physics}
The advent of the synchrocylotron made possible the study of `medium energy' nuclear physics in the 1950s and 60s. The energy range is similar to that of proton radiotherapy, 70 to 300\,MeV, so this is the nuclear physics relevant to the halo. Because binding energies of nucleons in nuclei are of order 10\,MeV, the incident proton easily ejects protons, neutrons and nucleon clusters from the target, but pion production is infrequent, and excited states of the proton itself cannot be reached. 

There were some 15 synchrocyclotrons world-wide, and this was an active field of study. A contemporary monogram by Clegg \cite{Clegg1965} lists 113 papers. We will discuss only those aspects relevant to the halo.

Medium energy reactions can be {\em coherent} (the proton interacts with the nucleus as a whole) or {\em incoherent} (the proton interacts with constituents of the nucleus). In medical physics terms these correspond respectively to elastic and inelastic reactions (the recoiling nuclide is the same as the target, but may be excited) and nonelastic reactions (the recoiling nuclide is different, and is frequently excited) \cite{icru63}.

Unlike ground state properties which, owing to the nuclear shell model, change dramatically from one nuclide to the next, medium energy effects vary rather smoothly with atomic number. Thus the probability of knocking a proton out of C is not greatly different from knocking one out of O at the same energy. This becomes relevant if, for instance, we substitute SolidWater (Gammex, Inc., 7600 Discovery Drive, Middleton, WI 53562, USA) for water in a halo experiment.

The link between experiment and proton therapy is the MC nuclear model, actually a collection of models which encapsulate the wide range of experimental data and theory just alluded to. Incoherent reactions, for instance, are first described by the `intranuclear cascade' Bertini model \cite{bertini}. The subsequent decay of the recoiling nucleus is handled by a separate model.

\subsection{Coherent reactions}\label{sec:coherent}
\begin{figure}[p]
\centering\includegraphics[width=2.91in,height=3.5in]{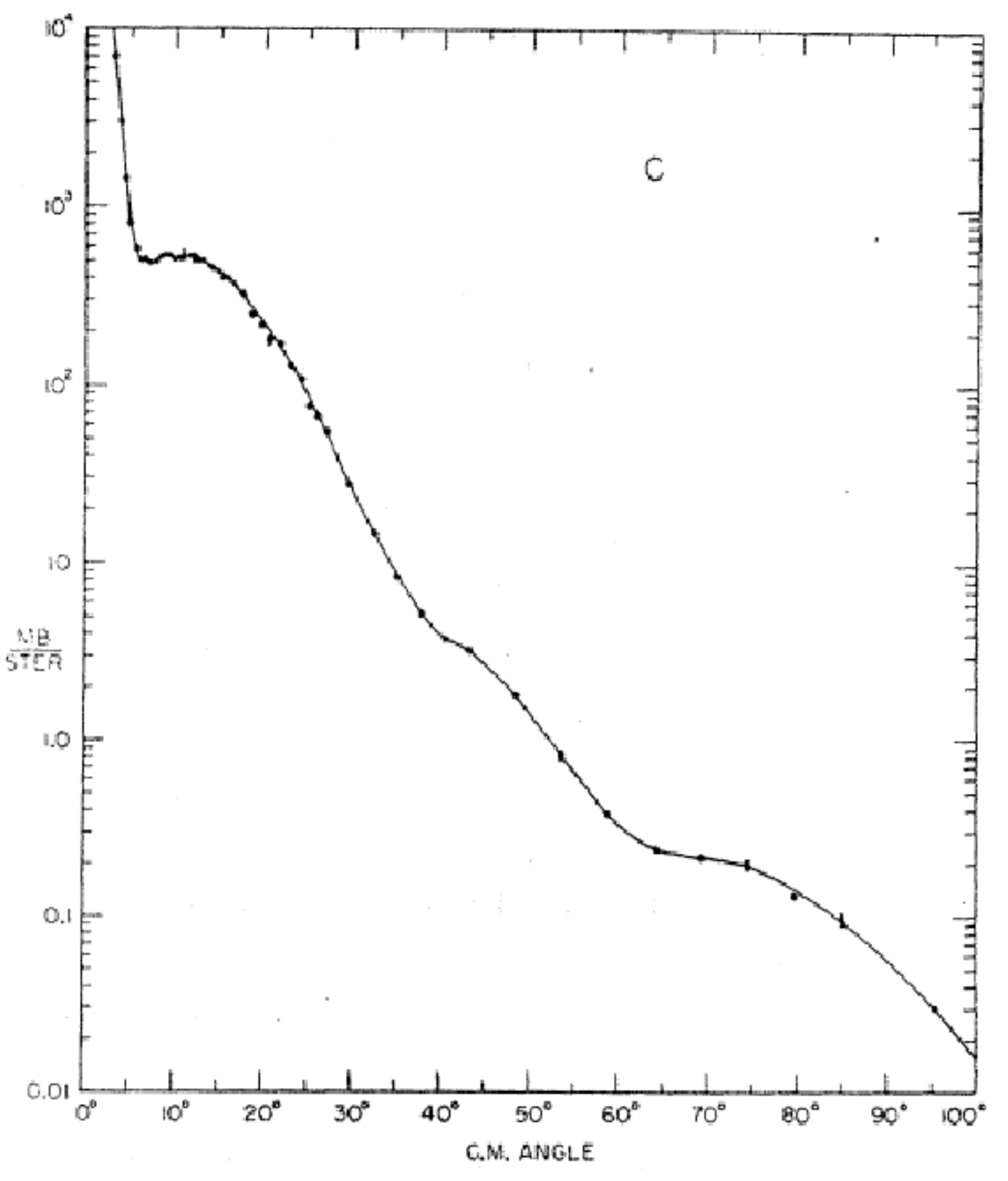} % 0.83
\caption{Gerstein et al. \cite{Gerstein1957} Fig.\,3: Elastic scattering differential cross section of 96\,MeV protons on carbon. Note (from left) the Moli\`ere single scattering tail, the Coulomb interference dip near 7\,\degr\ and scattering to large angles, with shallow diffraction minima, by the nuclear force. \label{fig:Gerstein3}}
\end{figure}

Of interest to us are elastic and inelastic scattering on O, $^{16}$O(p,p)$^{16}$O and $^{16}$O(p,p)$^{16}$O$^*$. (We include elastic scattering on free hydrogen $^1$H(p,p)$^1$H with the incoherent reactions for reasons explained below.) Experiments on O are scarce because of the difficulty of making thin O targets, but the reaction on C is similar. Fig.\,\ref{fig:Gerstein3}, taken from Gerstein et al. \cite{Gerstein1957}, illustrates the main points. It shows the differential scattering cross section as a function of center-of-mass (CM) angle which, for scattering on C, is nearly equal to the lab angle.\footnote{~The transformation from lab to CM is given in \cite{Gottschalk2014}.} Recalling that elastic scattering occurs anywhere along the path of a 177\,MeV incident proton, the 96\,MeV of Fig.\,\ref{fig:Gerstein3} is representative.  

The steep fall to 7\,\degr\ is the single scattering tail of the Moli\`ere distribution, coherent scattering on C via the Coulomb force. The more gradual drop thereafter is coherent scattering via the nuclear force. The `Coulomb interference' dip at 7\,\degr\ is due to quantum mechanical interference between the two processes, which have indistinguishable final states.

The entire pattern, including the barely resolved diffraction minima at larger angles, is well explained for all nuclides by an optical model of elastic scattering, which treats the nucleus as an absorbing disk with a diffuse edge. The pattern would be slightly narrower for O, with the Coulomb interference dip less pronounced and the diffraction minima more pronounced.

The takeaway message for the halo is that elastic scattering falls very rapidly at first, then much less so with an angular distribution more nearly exponential than Gaussian. Because the kinetic energy of the recoiling nucleus is small, an incident proton scattered elastically anywhere along its trajectory will still stop at nearly the same depth as an unscattered proton. 

All this explains the persistence of the Bragg peak to rather large distances from the beam, much larger than can be attributed to Moli\`ere single scattering. However, we may expect the resulting Bragg peak to pull forward at larger deflections because the projected depth is less than the pathlength and the kinetic energy of the recoiling nucleus increases with scattering angle.

\subsection{Incoherent reactions}
Common medium-energy incoherent reactions are proton knockout $^{16}$O(p,2p)$^{15}$N, $Q=-12.1$\,MeV and neutron knockout $^{16}$O(p,pn)$^{15}$O, $Q=-15.7$\,MeV. ($Q$ equals the rest energy of the reactants minus the rest energy of the products. If $Q$ is negative, kinetic energy is lost in the reaction.) These reactions obey a `quasi-elastic' (QE) or `quasi-free' model. They resemble free (p,p) or (p,n) scattering except that the target nucleon is {\em bound} (energy, at least $-Q$, is required to eject it) and {\em moving}. The motion corresponds to the `Fermi momentum' of the target nucleon or, more accurately for light nuclei, to the momentum distribution of appropriate states of the nuclear shell model.

Focusing for the moment on the (p,2p) reaction, by measuring the total kinetic energy of the outgoing protons one can measure the binding energy of protons in each shell, and from their angular distributions one can infer the momentum distributions. See  \cite{Gottschalk2014} for more details and \cite{Tyren1966} for the best (p,2p) measurement on O.  

\subsection{Size of the halo}\label{sec:size}
Incoherent reactions cause the broad dose enhancement in the halo at mid-range. We can estimate its radius using energy and momentum conservation. We only sketch the argument here; see \cite{Gottschalk2014} for mathematical details.

We are interested in the longest range and therefore the lightest charged secondaries, protons. (Helpfully, they are also the most common.) Also, we want reactions that divide the available energy among the fewest secondaries. That leads us to the $^1$H(p,p)$^1$H and $^{16}$O(p,2p)$^{15}$N reactions.

The first, elastic scattering on free H, has a 2-body final state (6 parameters). Energy and momentum conservation supply four equations, so if we assume the direction of either final proton (2 parameters) we can find its energy, therefore range, and everything about the other proton. 

The (p,2p) reaction has a 3-body final state and the target proton has some {\em a priori} unknown binding energy. Thus we have four additional unknowns. However, the binding energy and most likely nuclear recoil momentum (4 parameters) are known from (p,2p) experiments e.g. \cite{Tyren1966}, so we can then proceed as in the (p,p) case. 

\begin{figure}[p]
\centering\includegraphics[width=4.57in,height=3.5in]{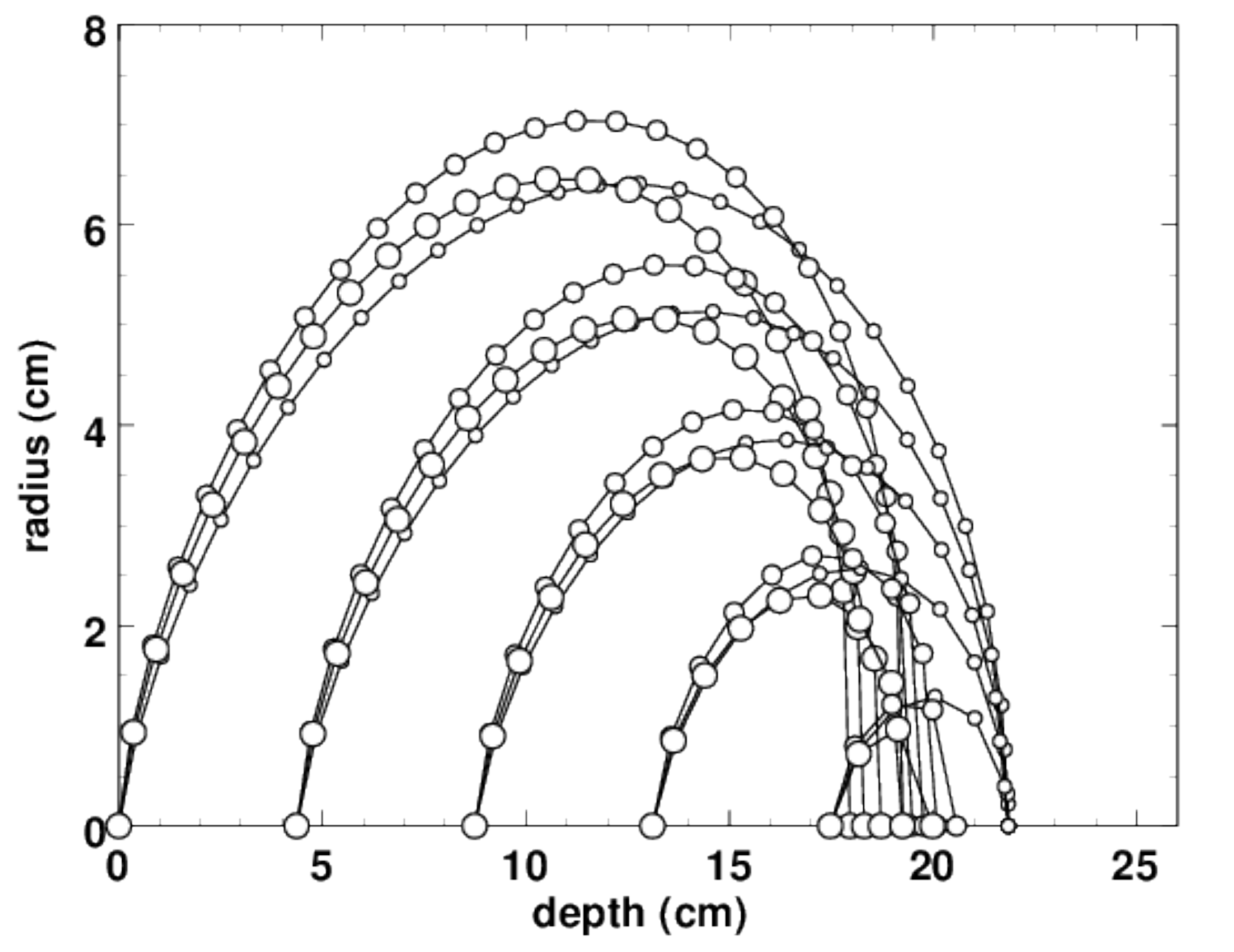}% 0.765
\caption{Outgoing proton stopping points in water for a 180\;MeV beam with reactions at five depths. Small, medium, large circles represent elastic, QE with $E_\mathrm{B}=12.4$\,MeV and QE with $E_\mathrm{B}=19.0$\,MeV. Recoil nucleus parameters are $\theta=300$\degr, $pc=75$\,MeV \cite{Tyren1966}.\label{fig:haloBump}}
\end{figure}

Putting all that together, we can compute the locus of most likely stopping points of the more energetic final proton assuming either a  $^1$H(p,p)$^1$H or $^{16}$O(p,2p)$^{15}$N reaction takes place at a given depth. Fig.\,\ref{fig:haloBump} is a plot of the results for reactions with parameters given in the caption and justified by \cite{Gottschalk2014,Tyren1966}. Because these only represent most likely values, and the core itself has some size, we must imagine a fairly broad band of maximum radii. Still, Fig.\,\ref{fig:haloBump} strongly suggests that a bump will be found around midrange in depth-dose scans far from the beam, in agreement with previous measurements \cite{Lin2013,Lin2014} and the present one.

If we do the same calculation at several incident energies and note the maximum radius at each we conclude that the radius of the halo is proportional to the range of the beam with a coefficient very roughly 1/3.

\section{Measuring the Halo}\label{sec:measurement}

\subsection{Purpose}
The goal was a comprehensive, absolute measurement of the entire halo, including the core and the transition to the aura, with an ADCL calibrated IC at a typical PBS energy. We chose 177\,MeV to compare with the previous work of the PSI group \cite{pedroniPencil}. 

\subsection{Method}
Fig.\,\ref{fig:haloSetup} shows the experimental setup in the Burr Center experimental beam line. Emerging from a 6.35\,cm diameter beam pipe through a 0.008\,cm Kapton vacuum window, the beam passed through a 150\,cm helium bag, a 30\,cm air gap and the beam monitor before entering water through a 0.86\,cm PMMA tank wall.

The beam monitor was an air filled PPIC of in-house design having five Al foils of total thickness 0.0064\,cm, placed immediately in front of the water tank to minimize spray. Its gain had previously been determined as 96.8 at 177\,MeV. Though hermetically sealed, it has slack windows and therefore behaves as a vented chamber.  Recombination has been shown to be negligible under our beam conditions. The electronics chain, also in-house, consisted of a 10\,pC/count recycling integrator \cite{charge}, a divide-by-10$^4$ prescaler, and a presettable counter that turned off the beam. The nominal calibration was $6.448\times10^9$\,p/monitor unit (MU).

A CRS (Computerized Radiation Scanners Inc., 140 Sopwith Drive, Vero Beach, FL 32968, USA) water tank was used for positioning only. The beam was offset to one side of the tank. A lateral extension to the ionization chamber holder allowed us to position the dosimeter at distances up to 10\,cm from nominal beam center. The dosimeter was an Exradin (Radiation Products Design Inc., 5218 Barthel Industrial Drive, Albertville, MN 55301) T1 biased to 300\,V, oriented with the stem vertical, and read out by a Keithley Model 6512 electrometer. The T1 ADCL calibration factor was 76.37\,cGy/nC $\pm1.4$\,\% at STP.

The T1 was positioned at distances from nominal beam center (`radii' $r$, see Sec.\,\ref{sec:symmetry}) ranging from 0 to 10\,cm and measurements were taken at discrete depths between 1.53\,cm water equivalent (the lowest obtainable) and 25\,cm (well beyond the Bragg peak). At each $r$ the number of MU was adjusted (between 2 and 500) to obtain a convenient electrometer reading, and from time to time the beam current was adjusted (between 1 and 33\,nA) to obtain a convenient counting time ($<1$\,s to 15\,s). The entire experiment including alignment took about 6 hours.
 
Two or more readings at each point, recorded in a long text file, were subsequently averaged and converted to absolute dose in MeV/g/p (1\,MeV/g = 0.1602\,nGy) by an in-house Fortran analysis, fitting and graphing program.

\subsection{Results}
\begin{figure}[p]
\centering\includegraphics[width=4.78in,height=3.5in]{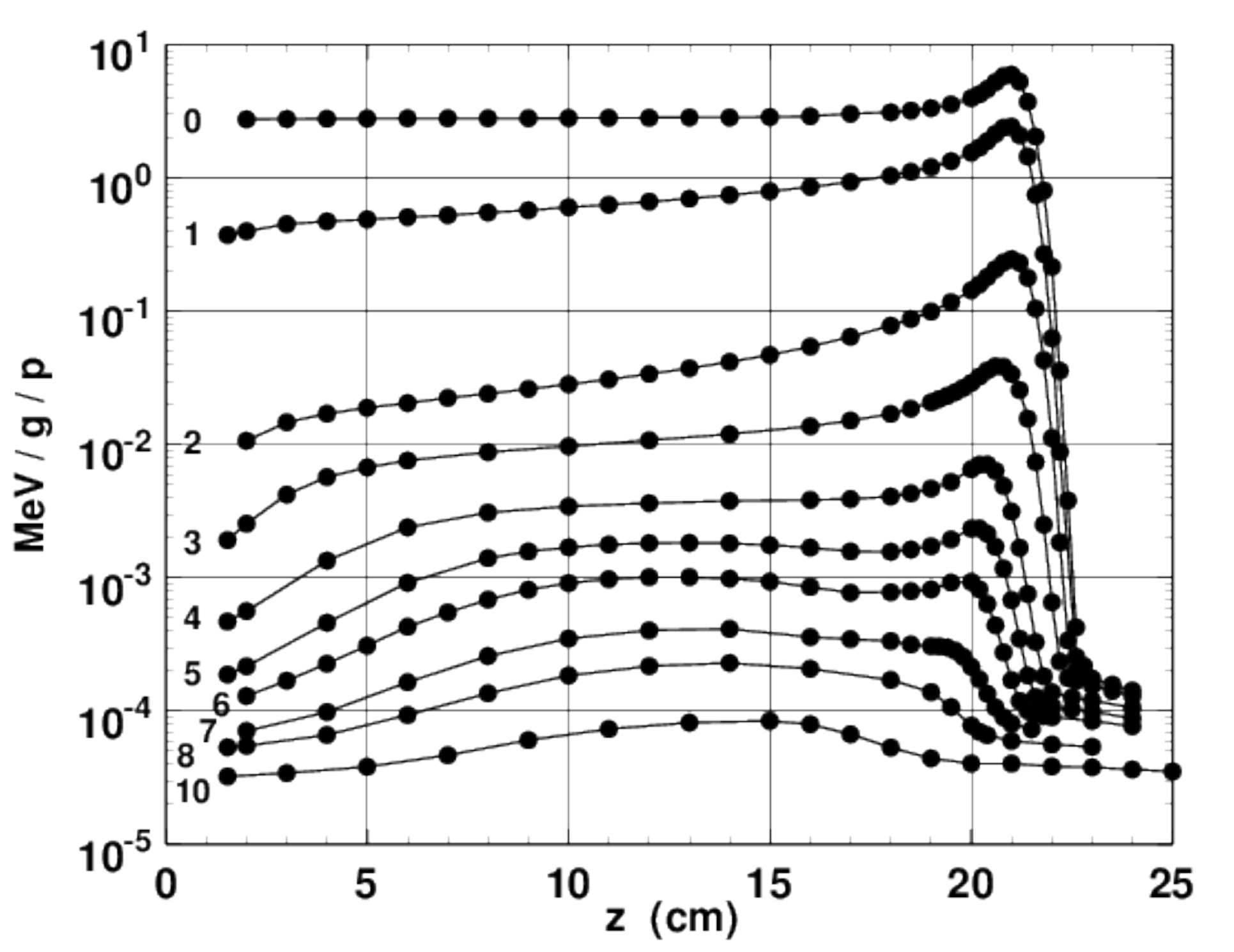}% 0.731
\caption{Doses measured in this experiment. Left-hand numbers are the distance (cm) of each depth scan from the beam center line. The lines serve only to guide the eye. 1\,MeV/g = 0.1602\,nGy.\label{fig:dmlg}}
\end{figure}

\begin{figure}[p]
\centering\includegraphics[width=4.66in,height=3.5in]{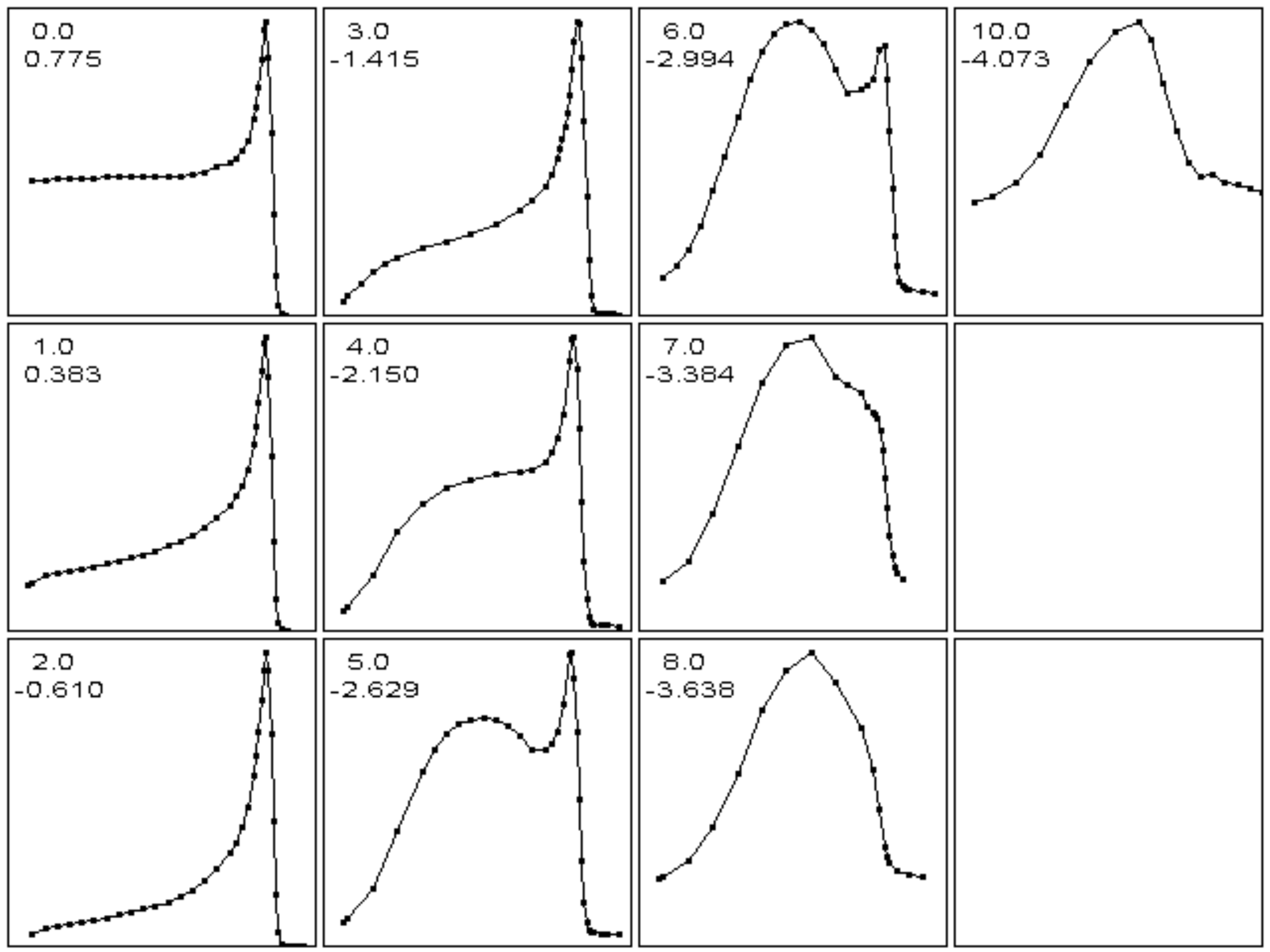}% 0.752
\caption{The same data as Fig.\,\ref{fig:dmlg} in auto normalized linear presentation. Upper number in each frame is $r$ (cm), lower is log$_{10}$(dose/(MeV/g/p)) for the highest point in the frame (cf. Table\,\ref{tbl:dmlg}).\label{fig:dmlin}}
\end{figure}

Table\,\ref{tbl:dmlg} gives our measurements (295 points) as log$_{10}$(D/(MeV/g/p)). Fig.\,\ref{fig:dmlg} shows them in semilog format, and Fig.\,\ref{fig:dmlin} in linear format. Fig.\,\ref{fig:dmlg} illustrates the virtue of longitudinal, rather than radial, scans and foretokens the difficulty of fitting any single radial profile to the data. A prominent feature we have already mentioned is the Bragg peak which persists to quite large radii, becoming somewhat shallower.  We also see the distinct change in character from the halo to the aura. In Fig.\,\ref{fig:dmlg} it shows up as an abrupt change in slope distal to the Bragg peak. In Fig.\,\ref{fig:dmlin} the aura is seen as a broad background to the midrange bump at the larger radii.

The transition from core to halo is marked by the very rapid (Gaussian) disappearance of the entrance dose with $r$ over the first three frames of Fig.\,\ref{fig:dmlin}. 

\subsection{Experimental issues}

\subsubsection{Absolute normalization}
Since both the monitor and field IC were effectively vented, it would have sufficed to know both calibrations at STP, independent of atmospheric pressure on the day of measurement, to compute the absolute dose per proton. Unfortunately the multiplication factor of 96.8 mentioned above was measured at an unrecorded pressure. (The test beam line is normally used for non-clinical measurements where a few percent absolute error is unimportant.) Before it could be recalibrated at known pressure, it was physically damaged and had to be rebuilt. The resulting uncertainty, say $\pm3$\,\%, dominates the absolute dose per proton.\footnote{~The standard deviation of 208K measurements over two years near sea level at Teddington, UK, is 1.05\%.}

\subsubsection{Beam alignment}
The dosimeter was initially aligned with a laser. Subsequently the beam was steered magnetically by $\approx\pm1$\,mm to maximize the signal when the dosimeter was at nominal $x=y=0$.

\subsubsection{Tank entrance wall}
The composition of the 0.86\,cm PMMA entrance wall was (8, 60, 32)\,\%  (H, C, O) by weight compared with (11, 89)\,\% (H, O) for water. Since medium energy cross sections change slowly with atomic number (Sec.\,\ref{sec:physics}), only an exceedingly careful experiment would be able to detect the non-water wall. (It should be noted that in some published measurements \cite{Lin2013,Lin2014} the {\em entire} stopping medium is SolidWater, which has more C and less O by weight than PMMA \cite{GammexSW}.)

\subsubsection{Detector size}
The active volume of the Exradin T1 thimble chamber in plan view is an annulus with inner and outer radii 0.5 and 2\,mm, for $\sigma_r=1.46$\,mm. Its length is 4.4\,mm, for $\sigma_y=1.27$\,mm. Subtracting the larger value in quadrature from the observed $\sigma_x=5.95$\,mm of the incident beam \cite{Gottschalk2014} gives $\sigma_{x,\,\mathrm{corr}}=5.77$\,mm, a 3\% effect which becomes smaller as the beam spreads with depth. We have ignored it. In principle a MC simulation could approximate the T1 geometry as accurately as desired.

\subsubsection{Leakage current}
Drift current was measured to be 0.5\,pC/15\,s just after the $r=10$\,cm (lowest dose rate) run. That is 2\% of the lowest signal recorded and we have ignored it. 

\subsubsection{Recombination in the dosimeter}
Recombination in the T1 was explored by taking data at the Bragg peak over a wide range of beam currents. The signal per MU changed less than 1\%.

\subsubsection{Positioning error}
Looking ahead at Sec.\,\ref{sec:fits} we see that two very different fits are low for $r=6$\,cm and high for the adjacent scans, suggesting positioning errors of order 1\,mm in $r$. Irregularities in the depth of the Bragg peak as a function of $r$ suggest an error of similar magnitude in $z$. We have not fitted or otherwise attempted to correct any such errors. 

\subsubsection{Symmetry}\label{sec:symmetry}
The beam entering the water tank had $\sigma_x=0.595$\,cm as determined by the MD fit to be discussed, and we found $\sigma_y=\sigma_x$ within a few percent using Gafchromic film. The size of the core has little effect on the halo \cite{Peeler2012}. $\sigma_x$ of the beam was 5.3$\times$ smaller than the beampipe radius, so there was no appreciable scraping, and the beam only passed through homogeneous objects (vacuum window, helium bag, air, beam monitor) after emerging from vacuum. 

Available space dictated a small water tank. The beam was offset to permit measurements at large $r$, and therefore was nearer the tank wall on the side opposite the dosimeter. It was also nearer the surface of the water than the bottom of the tank. Some charged reaction products directed away from the dosimeter will leave the water tank, but that will not affect the signal. There may be an exceedingly small missing dose from $\gamma$-rays and neutrons that would have backscattered into the dosimeter had the beam been surrounded by more water.

The underlying physics (Sec.\,\ref{sec:physics}) is cylindrically symmetric. Lin et al. \cite{Lin2014} confirm the symmetry of the 180\,MeV halo at various depths in Gammex SolidWater even when the incident beam is somewhat asymmetric.

All that leads us to assume that what we measured is one half of a symmetric distribution, and we discuss it accordingly. It would, of course, be desirable to confirm that experimentally in the future.

\subsubsection{Range of measurements}
The subsequent MI fit (see Sec.\,\ref{sec:MI}) required depth-doses spanning the same range in $z$, and we had to pad some of our scans with fabricated points. They are not shown in our graphs, and do not appreciably affect the fit in the region actually measured. However, this should be avoided in future measurements.

\section{Fitting the Halo}\label{sec:fits}
Any fit to a phenomenon as complicated as the core/halo (Fig.\,\ref{fig:dmlin}) will itself be complicated. Here we only try to give the flavor of the MD and MI fits, and refer the reader to \cite{Gottschalk2014} for fuller explanations, mathematical details, optimization technique, definitions and numerical values of parameters.

We have used Janni's 1982 values \cite{janni82} for the range-energy relation of protons in water. A growing body of experimental evidence \cite{moyers,Siiskonen2011,cascio07} suggests these are quite accurate and, in particular, better than ICRU Publication\,49 \cite{icru49} which lists ranges some 0.9\% smaller.

\subsection{Goodness-of-fit criterion}
\begin{figure}[p]
\centering\includegraphics[width=4.72in,height=3.5in]{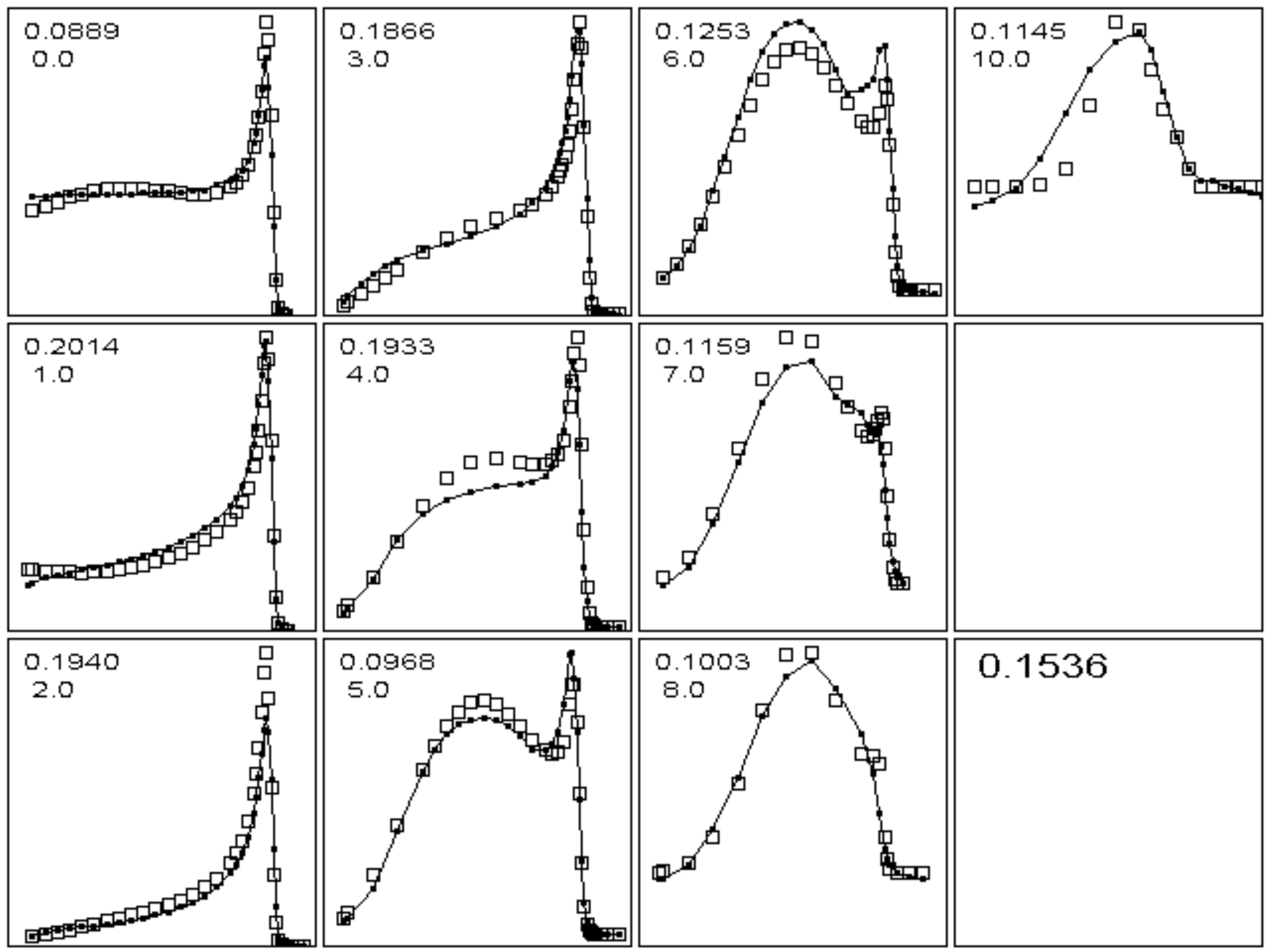}% 0.741
\caption{The same as Fig.\,\ref{fig:dmlin} with model-dependent (MD) fit of each measured point (empty square). The upper number in each frame is the goodness of fit, the rms value of (measurement/fit\,$-$\,1) for that frame. The isolated number is its overall value. \label{fig:dlinMD}}
\end{figure}

The MD and MI functions and optimization techniques are entirely different but the final figure of merit is the same. We seek a least-squares fit, but ordinary $\chi^2$ is not suitable because of the orders-of-magnitude variation of $D(r,z)$. ($D\equiv$\,physical absorbed dose, $r\equiv$\,radius, $z\equiv$\,water equivalent depth.) Therefore we minimize $\chi^2$ as computed in $\log_{10}$ space. If the fit is relatively good, that is equivalent to minimizing the mean squared deviation from 1 of the {\em ratio} of measured $D_M$ to fitted $D_F$ since, provided $D_{M}/D_{F}$ is not too different from 1, it is easy to show that 
\begin{equation}\label{eqn:fom}
\log_e10\,\left(\sum_{i=1}^N\Big(\log_{10}D_{Mi}-\log_{10}D_{Fi}\Big)^2\right)^{1/2}\,\approx\,\left(\sum_{i=1}^N\Big(\frac{D_{Mi}}{D_{Fi}}-1\Big)^2\right)^{1/2}
\end{equation}
It is the quantity defined by the LHS of Eq.\,(\ref{eqn:fom}) that is given in each frame of Fig.\,\ref{fig:dlinMD}\;.

\subsection{Model-Dependent fit}\label{sec:MD} 
Physical doses add, so the dose from a pencil beam in a large water tank can be written
\begin{equation}\label{eqn:D}
D(\vec x)\,=\,D(r,z)\,=\,D_\mathrm{em}(r,z)\,+\,D_\mathrm{el}(r,z)\,+\,D_\mathrm{ne}(r,z)\,+\,D_\mathrm{au}(r,z)
\end{equation}
The radial symmetry we have assumed here obtains for the core if the incident beam is radially symmetric (as ours was) but will be nearly true for the other terms even if it is not \cite{Peeler2012}. It is not a fundamental assumption. The fit could be generalized to $D(x,y,z)$ with a little more work.

Fig.\,\ref{fig:dmlin} exhibits four superimposed patterns of dose deposition which require four mathe\-mati\-cal descriptions, hence the four terms in Eq.\,(\ref{eqn:D}). The subscripts conform to the dominant process at work in each: multiple EM, elastic (coherent), nonelastic (incoherent) and aura (neutrals). $D_\mathrm{em}$ corresponds to the core and $D_\mathrm{el}+D_\mathrm{ne}$ to the halo.
 
$D_\mathrm{em}$ includes energy deposited by recoil nuclei, whose spatial distribution mimics the core because of their short range. This is the `nuclear stopping power' discussed in \cite{icru49}, whose relative contribution to $D_\mathrm{em}$ is tiny, 0.04\% in water at 100\,MeV. 

$D_\mathrm{el}$ includes {\em inelastically} scattered protons from O, whose decay gammas contribute to $D_\mathrm{au}$. $D_\mathrm{el}$ also includes elastically scattered protons from H if one of them happens to be forward (cf. Fig.\,\ref{fig:haloBump}). The Moli\`ere single scattering tail from H and O contributes to both $D_\mathrm{em}$ and $D_\mathrm{el}$ because of the smooth transition in Moli\`ere theory between the central Gaussian and the single scattering tail cf. \cite{transport2012} Fig.\,10\;. 

$D_\mathrm{ne}$ includes a host of reactions, usually written (p,px), in addition to the quasi-elastic (p,2p) and (p,pn) we have focused on. Both the scattered proton and the knocked-out nucleons and clusters contribute to the dose.

Without losing generality we can write dose as fluence $\Phi$ (p/cm$^2$), the areal density of particles at a point in space, times mass stopping power $S/\rho$ (MeV/g)/(p/cm$^2$), their average mass rate of energy deposition. Fluence in turn can be expressed as the product of a normalized function $F$ times a dimensionless normalizing constant $M$. Thus
\begin{equation}\label{eqn:MFS}
D(r,z)\,=\,M_\mathrm{em}F_\mathrm{em}(S/\rho)_\mathrm{em}\,+\,M_\mathrm{el}F_\mathrm{el}(S/\rho)_\mathrm{el}\,+\,\ldots
\end{equation}
That decomposition is useful only if we have theories for $F$ and/or $(S/\rho)$ for the process in question. That is completely true for `em', less so for `el' and so on. By the time we reach `au' the fit is purely empirical and we will simply fit $D_\mathrm{au}$ directly.

In sketching the MD fit, we emphasize the core, loosely thought of as `the pencil beam', because that is likely to survive as part of the final clinical parameterization. We remind the reader to consult \cite{Gottschalk2014}, especially Appendix\,D.3, for details. For clarity, we have suppressed computational details such as using dimensionless normalized values $r_\mathrm{n},z_\mathrm{n}$ instead of $r,z$.

\subsubsection{Core (9 parameters)}\label{sec:core}
Following Eq.\,(\ref{eqn:MFS}) the core may be written
\begin{equation}\label{eqn:Dem}
D_\mathrm{em}\,=\,(1-\alpha(z)\,z)\;\times\;G_\mathrm{2D}(r;\sigma_\mathrm{em}(z))\;\times\; S_\mathrm{em}(z_\mathrm{adj}(r);\sigma_\mathrm{sem}(r))\;/\;\rho_\mathrm{water}
\end{equation}

The first term is the primary attrition function, written so as to emphasize that it is nearly linear in $z$. $\alpha$ differs from the constant ($\approx0.01$/cm) employed by Bortfeld \cite{bortfeld}, which only accounts for {\em nonelastic} reactions \cite{janni82}. Our $\alpha$ is itself linear in $z$ (2 parameters).

The fluence describing the transverse profile is a normalized 2D (`cylindrical') Gaussian $G_\mathrm{2D}$ in $r$ whose $\sigma$ is a function of $z$ found by using FE theory. That requires the parameters of the incident beam: its range $R_0$ and three phase space parameters, the FE moments $A_0,A_1,A_2$ (4 parameters). If the core were not cylindrically symmetric we would require an additional three FE moments.

The stopping power $S_\mathrm{em}$ is a function of $z$ obtained by convolving the Bethe-Bloch EM stopping power (requiring $R_0$, already counted) with a 1D Gaussian in $z$ whose $\sigma_\mathrm{sem}$ combines range straggling with the energy (range) spread of the incident beam (1 parameter).

Later, we will use $S_\mathrm{em}$ off axis where the Bragg peak shifts and broadens. We introduce two more parameters defined, in the notation of \cite{Gottschalk2014}, by $z_\mathrm{adj}\equiv1+p_{12}\,r^2$ and $\sigma_\mathrm{sem}\equiv p_2+p_{13}\,r$\,. $p_2$ is already counted, so the grand total is 9 parameters. Five simply describe the incident beam and the last two are irrelevant for $r\approx0$. The only truly empirical parameters are the two coefficients describing the attrition of primaries.

\subsubsection{Elastic/Inelastic Term (4 parameters)}\label{sec:elastic}
Elastic scattering falls steeply for the first few degrees (Moli\`ere  single scattering) and exponentially thereafter (Fig.\,\ref{fig:Gerstein3}). Reactions may occur anywhere along the primary track but, because the angles are relatively small, the scattered protons stop at nearly the same depth. The elastic (coherent) term in the halo is 
\begin{equation}
D_\mathrm{el}\;=\;\alpha(z)\,z\,M_\mathrm{el}\;\times\;E_\mathrm{2D}(r;r_\mathrm{el}(r))\;\times\;
  S_\mathrm{em}(z_\mathrm{adj}(r);\sigma_\mathrm{sem}(r))\;/\;\rho_\mathrm{water}
\end{equation}
The first factor is justified because elastically scattered protons are secondaries by our definition. $M_\mathrm{el}$ is an empirical constant. $E_\mathrm{2D}$ is a 2D exponential function describing the fluence, with a variable slope that depends quadratically on $r$. Thus $D_\mathrm{el}$ introduces four new parameters. 

The raw data (Fig.\,\ref{fig:dmlg}) show that, at larger radii, the Bragg peak shifts upstream and broadens slightly. The shift is due to geometry (depth smaller than pathlength) plus the greater kinetic energy taken up by the recoiling nucleus at greater momentum transfer. The broadening is probably due to unresolved low lying nuclear states excited at greater momentum transfer. Rather than change the basic form of $S_\mathrm{em}$ we describe these small effects by introducing an adjusted $z_\mathrm{adj}(r)$ (quadratic in $r$) and $\sigma_\mathrm{sem}(r)$ (linear in $r$) which reduce to their simple values at $r=0$ and whose parameters were already introduced and counted in Sec.\,\ref{sec:core}.  

\subsubsection{Nonelastic Term (9 parameters)}\label{sec:nonelastic}
The nonelastic term, which produces the bump in dose near midrange at large radii, is
\begin{equation}
D_\mathrm{ne}\;=\;\alpha(z)\,z\,\,G_\mathrm{1D}(z;c(r),z_0(r),h(r))\;\times\;E_\mathrm{2D}(r;r_\mathrm{ne}(r))\;\times\;
  S_\mathrm{em}(z_\mathrm{adj}(r);\sigma_\mathrm{sem}(r))\;/\;\rho_\mathrm{water}
\end{equation}
The bump itself is a 1D Gaussian whose normalization $c$, mean value $z_0$ and halfwidth $h$ are each linear in $r$ (6 parameters). The transverse distribution (fluence) is exponential with its own variable slope, quadratic in $r$ (3 parameters). $S_\mathrm{em}/\rho_\mathrm{water}$ could probably be dispensed with, but does not add any parameters. It sets an absolute scale and (because of $G_\mathrm{1D}$) is only effective at midrange where it is nearly flat.

\subsubsection{Aura (3 parameters)}  
The aura, a slowly varying background, is represented by the entirely empirical
\begin{equation}
D_\mathrm{au}=p_{23}\;E_\mathrm{2D}(r;r_\mathrm{au}(r))
\end{equation}
where $r_\mathrm{au}$ is linear in $r$.

\subsubsection{Results}
Fig.\,\ref{fig:dlinMD} shows the MD fit (hollow squares) in linear presentation. The upper number in each corner equals the goodness-of-fit for that depth-dose, that is, the rms value of (measured/fit - 1) Eq.\,(\ref{eqn:fom}). The isolated number at lower right is the overall goodness-of-fit, approximately 15\%. The lower number in each corner is the radius at which that depth-dose was measured.

\begin{figure}[p]
\centering\includegraphics[width=4.72in,height=3.5in]{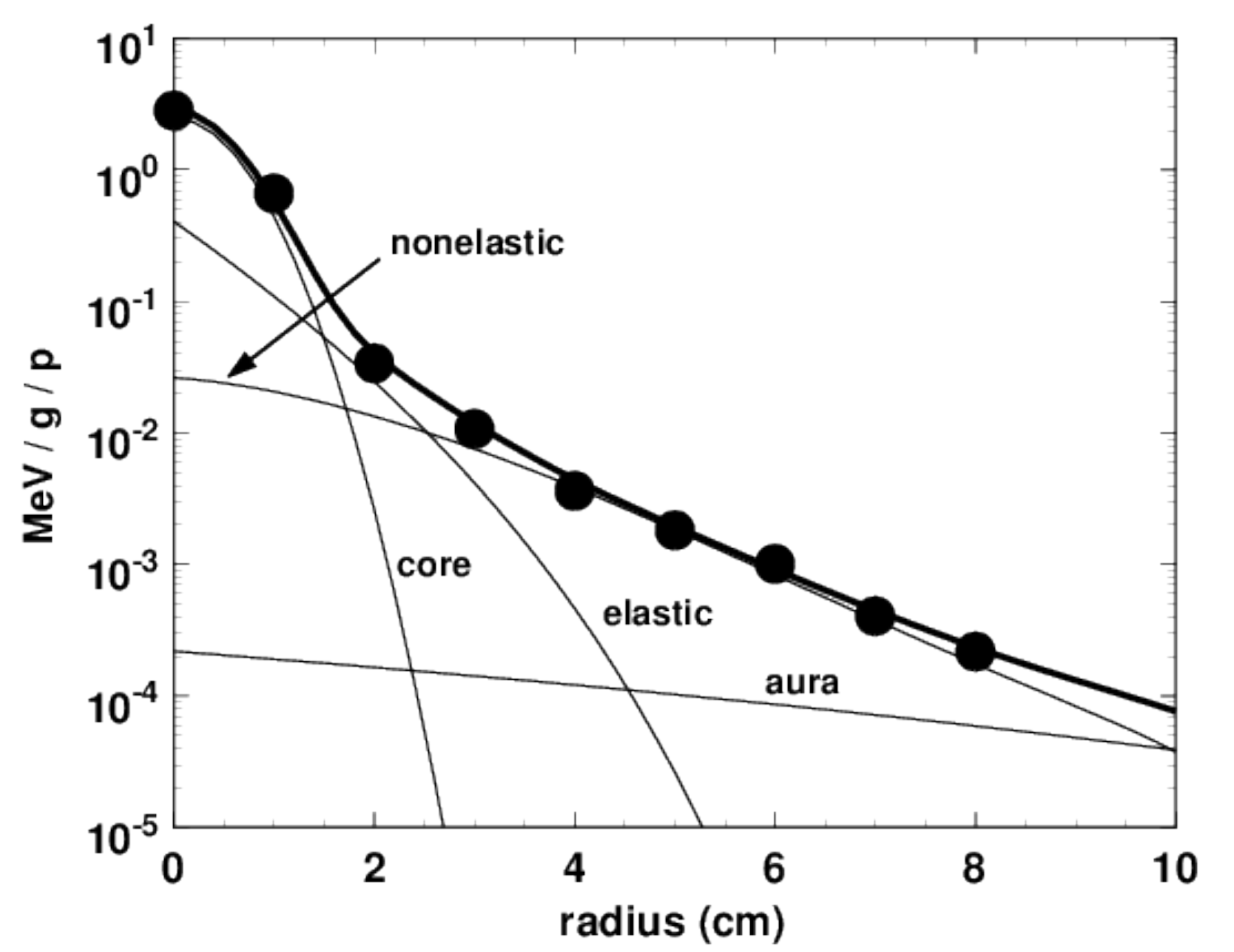}% 0.741
\caption{Bold line: MD fit to the transverse dose distribution at $z=12$\,cm (midrange) with experimental points (full circles). Light lines: contribution of each term in Eq.\,(\ref{eqn:D}).\label{fig:trans12}}
\end{figure}

\begin{figure}[p]
\centering\includegraphics[width=4.72in,height=3.5in]{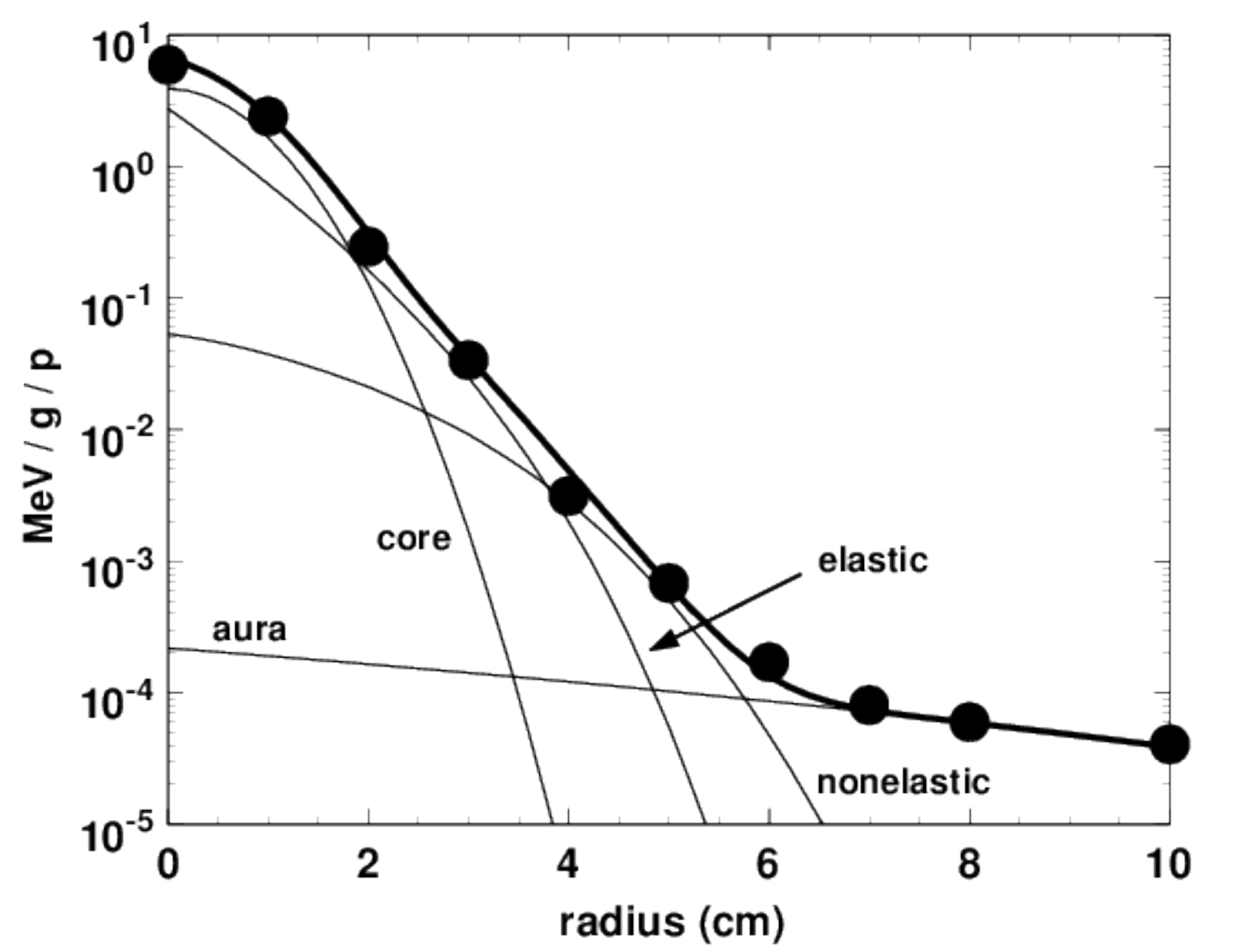}% 0.741
\caption{The same as Fig.\,\ref{fig:trans12} except $z=21$\,cm (end of range).\label{fig:trans21}}
\end{figure}

The MD fit allows us to plot separately the transverse variation of each term Eq.\,(\ref{eqn:D}) as shown in Fig.\,\ref{fig:trans12} (transverse scan at $z=12$\,cm, midrange) and Fig.\,\ref{fig:trans21} (transverse scan at $z=21$\,cm, near end-of-range). These figures illustrate the difficulty of characterizing the transverse distribution by any single analytic function.

\subsection{Model-Independent fit}\label{sec:MI}

\subsubsection{Method}
For a model-independent (MI) fit it makes sense to fit the logarithm of dose (Fig.\,\ref{fig:dmlg}) rather than dose itself. We tried and failed to fit log$_{10}D(r_j,z_{ij})$ with a 2D polynomial (a polynomial in $r$ whose coefficients are polynomials in $z$ or vice versa). The functions shown in Fig.\,\ref{fig:dmlg} are not well fit by polynomials of any reasonable order. 

Eventually we performed cubic spline fits to each curve of Fig\;\ref{fig:dmlg}. That yields 10 continuous functions of $z$, which can be used to create a virtual radial scan at any $z$ between 2 and 25\,cm, which can in turn be fit by a cubic spline, imposing zero slope at $r=0$ because the dose there is dominated by the core Gaussian.\footnote{~The `natural' boundary condition \cite{nr} is used elsewhere.} By this means a $101\times231$ matrix of log$_{10}$($D$/(MeV/g/p)) values on a 0.1\,cm grid spanning 0\,-\,10\,cm in $r$ and 2\,-\,25\,cm in $z$ is computed and written to a text file, which can at a later time be read and used with bilinear interpolation to obtain a continuous $D(r,z)$ \cite{Gottschalk2014}.

Note that we use spline {\em fitting} (regression) not {\em interpolation} (a more common application) throughout. That averages out any noise in the measurements. Because cubic spline regression, in particular the automatic selection of initial spline points, is far from standard, we describe our method fully in \cite{Gottschalk2014}. Spline fits extrapolate gracefully outside the fitted range, emphatically not true of high order polynomials. 

\subsubsection{Results}\label{sec:MIresults}
\begin{figure}[p]
\centering\includegraphics[width=4.72in,height=3.5in]{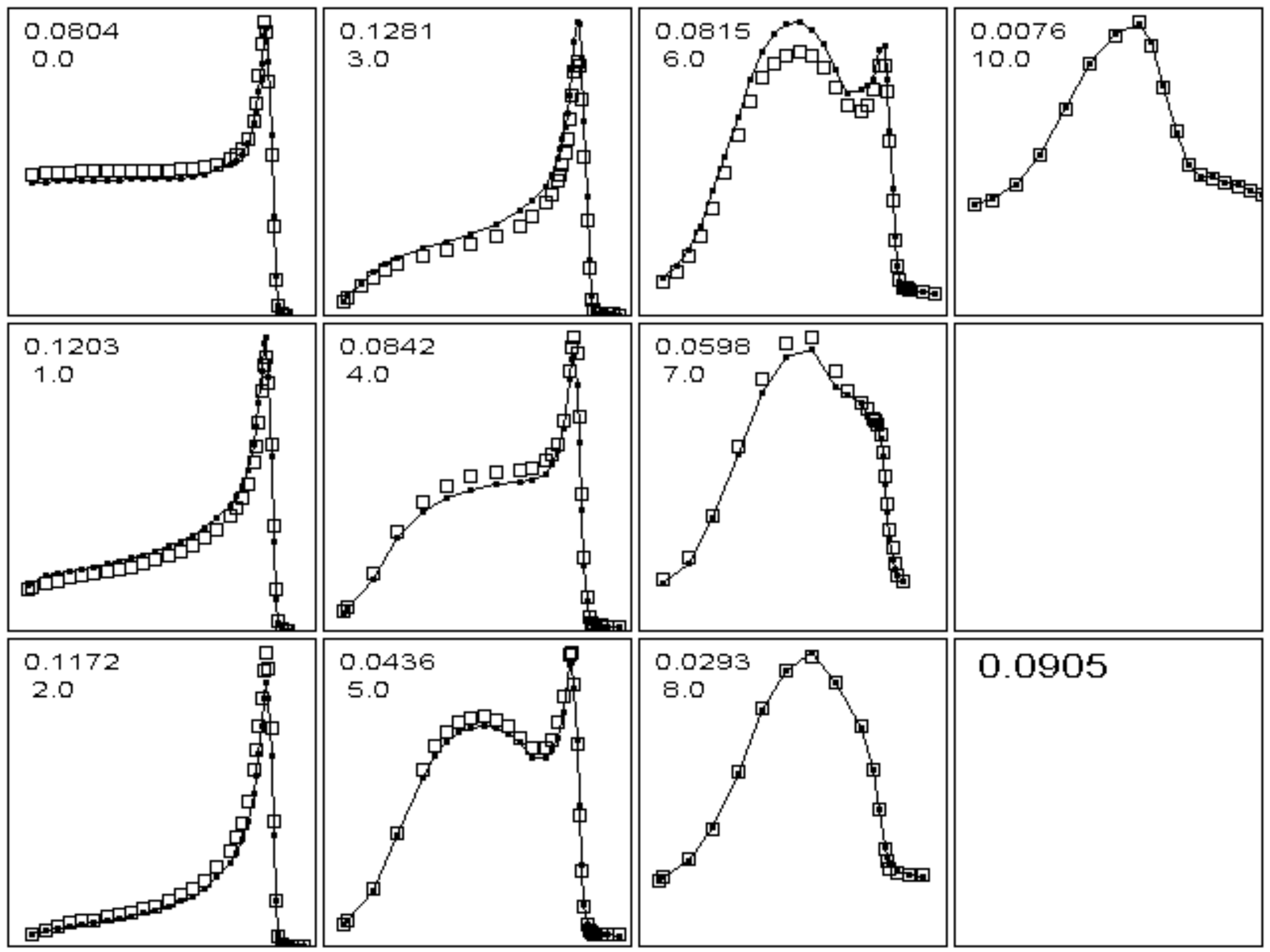}% 0.741
\caption{The same as Fig.\,\ref{fig:dlinMD} except empty squares show the model-independent (MI) fit.\label{fig:dlinMI}}
\end{figure}

\begin{figure}[p]
\centering\includegraphics[width=4.57in,height=3.5in]{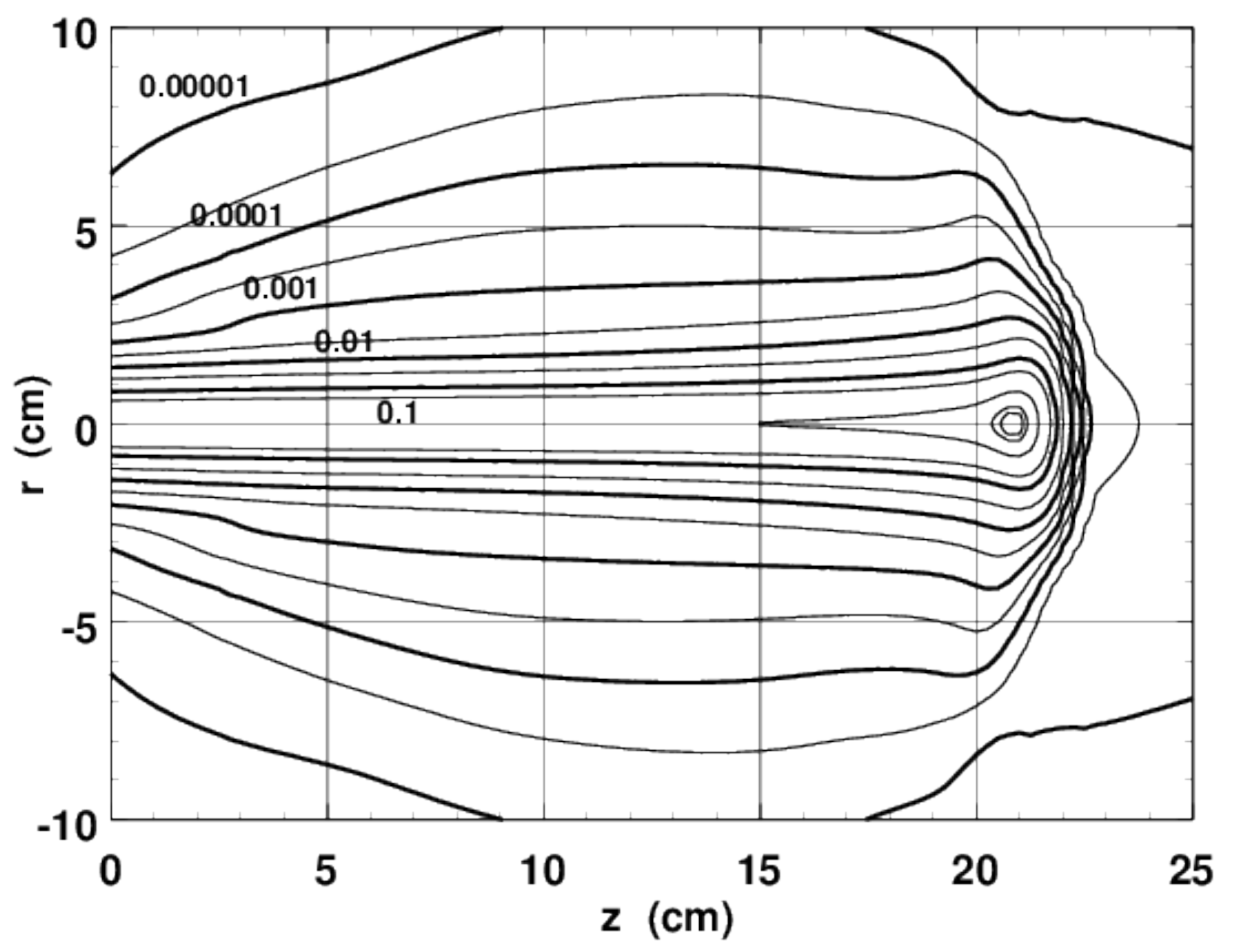}% 0.766
\caption{Contours of the MI fit at .9 .8 .5 .2 .1 .0316 .01 $\ldots$ 0.00001 of maximum dose. For visual effect we have treated the dose distribution as symmetric. Only the lower half was actually measured.\label{fig:DMIcontours}}
\end{figure}

Fig.\,\ref{fig:dlinMI} presents the MI fit in similar fashion to Fig.\,\ref{fig:dlinMD}. The overall compliance is a somewhat improved 9\%.  Fig.\,\ref{fig:DMIcontours} is a contour plot of the MI fit. Note that the labeled contours vary by decades, and that we have, for visual effect, assumed that our measurements represent one half of a cylindrically symmetric distribution.

\subsection{PSI fit}
A review of the literature \cite{Gottschalk2014} shows that the PSI fit has been used universally until now. Therefore their paper \cite{pedroniPencil} merits special attention. Here we simply compare the PSI fit (adjusted to account for our larger beam) with our data, leaving a conceptual discussion for the next section.
 
\begin{figure}[p]
\centering\includegraphics[width=4.70in,height=3.5in]{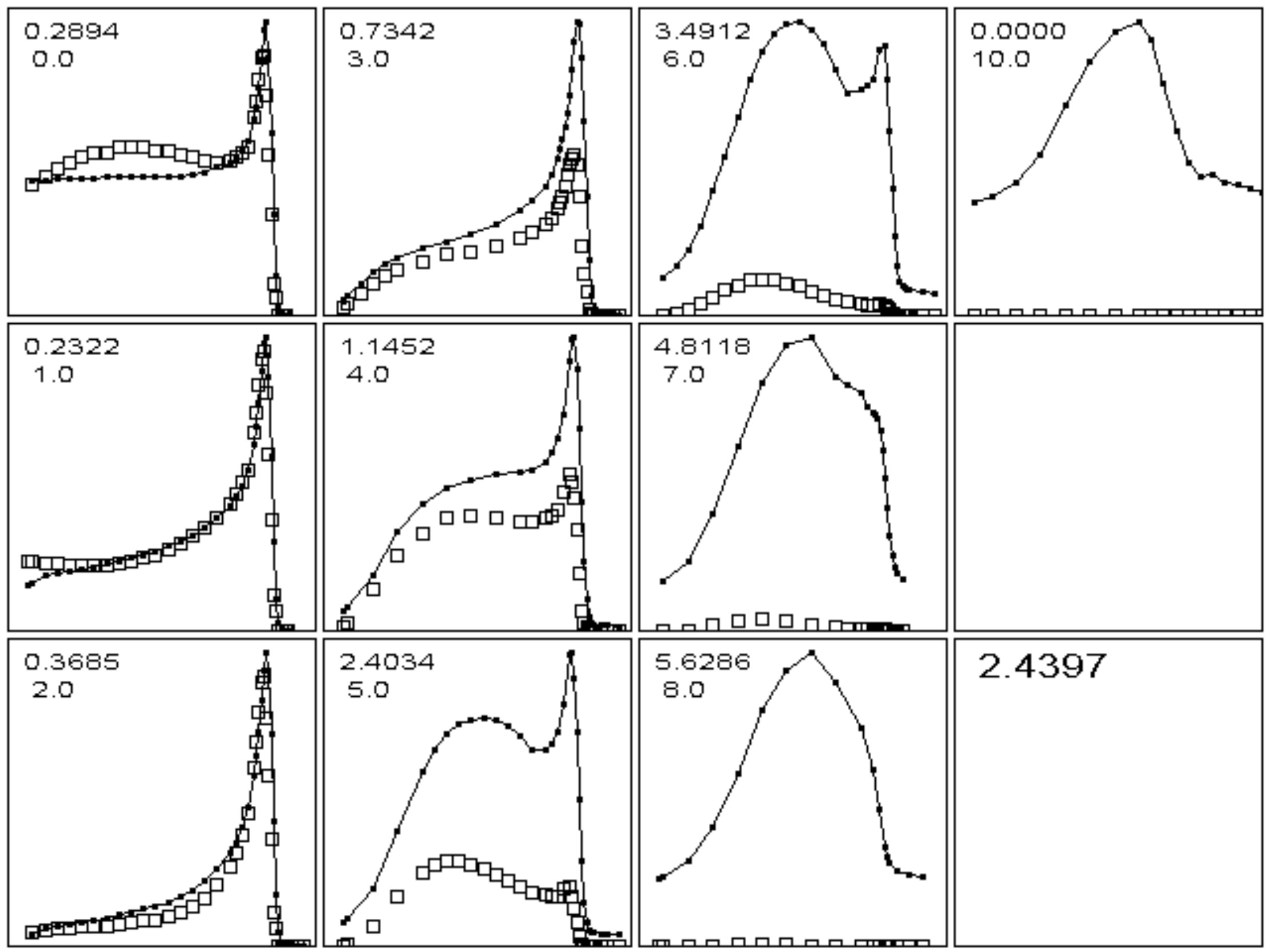}% 0.744
\caption{The same as Fig.\,\ref{fig:dlinMD} except the empty squares show the PSI fit adjusted by using our beam size $\sigma_\mathrm{em}(z)$ for $\sigma_\mathrm{P}(w)$ (no other changes, absolute comparison). Note the excess dose at midrange for $r=0$.\label{fig:PedLinFitEM}}
\end{figure}

We distilled their fit at 177\,MeV from their Figs.\,1, 3, 6 and 7 and checked our work by reproducing their Fig.\,5. Then we substituted our $\sigma_\mathrm{em}(z)$ (Sec.\,\ref{sec:core}) for theirs because our beam was larger. Fig.\,\ref{fig:PedLinFitEM} shows our data with the adjusted PSI fit.\footnote{~ The unadjusted PSI fit is shown in Fig.\,17 of \cite{Gottschalk2014}.} The comparison is absolute; the fit has not been normalized to the data.

The most obvious problem is the Gaussian transverse fluence of the halo, which falls far too rapidly with radius. In justice to \cite{pedroniPencil} the authors themselves described that as a `first preliminary estimate'. Others \cite{Fuchs2012,Knutson,Li2012} have improved on it using Cauchy-Lorentz or Voigt functions. Our experiment shows, however, that the transverse distribution depends rather strongly on depth and therefore, that no single transverse form characterizes the entire halo.

The PSI fit in the Bragg peak region at small $r$ is comparable to our MI fit and better than our MD fit. This region is sensitive to the Moli\`ere single scattering tail, and the transverse profile of our MD fit obviously needs improvement.

Let us focus, however, on the dose excess of the PSI fit around midrange at $r=0$. That is also in the high-dose region and therefore clinically significant. We now argue that this excess (not evident in either of our fits) may be due to the use of an inappropriate stopping power in the core term.

\section{Discussion}\label{sec:discussion}

\subsection{Comments on the PSI Fit}
The basic structure of the PSI fit \cite{pedroniPencil} is set by their Eq.\,(7) (their $w$ is our $z$) 
\begin{equation}\nonumber
D(x,y,w)\;=\;T(w)\times\left[\,(1-f_\mathrm{NI}(w))\times G_2^p(x,y,\sigma_\mathrm{P}(w))\;+\;f_\mathrm{NI}(w)\times G_2^\mathrm{NI}(x,y,\sigma_\mathrm{NI}(w))\,\right]\quad\hbox{(PSI\;7)}
\end{equation}
Let us focus on the first term, similar to our Eq.\,(\ref{eqn:Dem}) except for $T(w)$ in place of our $S_\mathrm{em}(r)/\rho_\mathrm{water}$\,. $T(w)$ is `$\ldots$ the total dose (in Gy\,cm$^2$) integrated over the whole plane perpendicular to the beam at the depth $w$.' tantamount to
\begin{equation}\label{eqn:Smixed}
T(z)\;\equiv\;\int_0^{2\pi}\int_0^{r_\mathrm{halo}}D(r,z)\;r\,dr\,d\phi\;\equiv\;S_\mathrm{mixed}(z)/\rho_\mathrm{water}
\end{equation}
Dimensionally, and by its role in Eq.\,(PSI\;7), $T(w)$ is a {\em mass stopping power} which we will denote $S_\mathrm{mixed}(z)/\rho_\mathrm{water}$. It describes a {\em mix} of primaries and secondaries. Let us explore the implications of using $S_\mathrm{mixed}$ instead of our $S_\mathrm{em}$. 

Note first that $S_\mathrm{mixed}$ can be measured, but not calculated from first principles, whereas $S_\mathrm{em}$ by contrast can be calculated from first principles, but not directly measured. (No particle is known to exist with the mass and charge of a proton but no nuclear interactions. The muon is closest.) 

Per Eq.\,\ref{eqn:Smixed}, $S_\mathrm{mixed}$ can be measured in a water tank with a BPC and a single pencil beam (Fig.\,\ref{fig:depthScans} top). That requires a sufficiently large BPC. It can also be measured with a small IC in a sufficiently broad uniform beam (Fig.\,\ref{fig:depthScans} bottom). Appendix\,\ref{sec:depthScans} proves the equivalence of the two methods. Both types of measurement are reported in \cite{pedroniPencil} and appear to be consistent although, as discovered later, their 4\,cm radius BPC was somewhat too small. Their Fig.\,1 reports BPC measurements at seven energies,\footnote{~The caption should read `per proton', not 'per 10$^6$ protons'.} whereas the topmost curve of their Fig.\,5 is essentially a broad-beam, small IC, measurement of $S_\mathrm{mixed}$ at 177\,MeV.

Even though $S_\mathrm{em}$ cannot be measured directly, it is easy enough to calculate \cite{Gottschalk2014}. The reason we favor it over $S_\mathrm{mixed}$ for the core is simple. Protons that {\em escape} hard scatters lose energy according to the Bethe-Bloch formula, that is, according to $S_\mathrm{em}$. In fact, \cite{pedroniPencil} itself states (emphasis ours) 
\begin{quote}
`The energy delivered by the {\em undisturbed protons} at a given depth $w$ is then obtained by the $dE/dx(w)$ value multiplied by the residual proton flux $I(w)$ at that depth.'
\end{quote}
making their use of $S_\mathrm{mixed}$ for the first term in Eq.\,(PSI\;7) all the more surprising. That choice has both numerical and operational consequences, which we discuss in turn.

\subsection{Numerical difference between $S_\mathrm{mixed}$ and $S_\mathrm{em}$}
$S_\mathrm{mixed}$ conflates the stopping power of secondaries and primaries (Fig.\,\ref{fig:depthScans} top).  At any given $z$ the secondaries, with lower energies, have greater stopping power. The difference is greatest near midrange. Near full range we only have primaries or (nearly the same) elastically scattered protons, and the difference vanishes.

To see the difference at 160\,MeV we need go no further than Fig.\,2 of  \cite{pedroniPencil}. Curve (a) is $S_\mathrm{mixed}$ and curve (e), labeled `dose from primary protons' (n.b.) is $S_\mathrm{em}$ times the attrition function $I(w)$. The difference at midrange is $\approx13\%$.
 
The difference can also be found from our data as shown in Fig.\,\ref{fig:S_EM}. $S_\mathrm{mixed}$ is simply the integral of our MD fit over radius cf. Eq.\,\ref{eqn:Smixed}, and $S_\mathrm{em}$ can be obtained from the fit itself. Because  $S_\mathrm{mixed}$ inextricably includes the primary attrition function ($I(w)$ in \cite{pedroniPencil}), we have multiplied $S_\mathrm{em}(z)$ by $(1-cz)$ adjusting $c$ to equalize the peak values. The excess in $S_\mathrm{mixed}(z)$ near midrange is then $\approx10\%$.

(Incidentally, the fact that $T(w)$ includes $I(w)$ raises another red flag. The function $(1-f_\mathrm{NI}(w))$, which seems to have the same general purpose as $I(w)$, is explicit in Eq.\,(PSI\;7). Therefore the $I(w)$ contained in $T(w)$ is duplicative. Our $S_\mathrm{em}$ does not include a primary attrition function.) 

Fig.\,\ref{fig:PedLinFitEM} shows a 20\% dose excess on axis at midrange for the PSI fit at 177\,MeV compared to our data. That cannot be interpreted solely in terms of the stopping power difference because there are also empirical functions $f_\mathrm{NI}(w)$ and $\sigma_\mathrm{NI}(w)$ in the fit. However, it is of the expected sign and order of magnitude.

We do not imply that any TPS now in use has dose errors of that magnitude. We do contend that using $S_\mathrm{mixed}$ introduces an excess of order 10\% in the central dose of a single pencil beam at midrange, and that is probably compensated in practice by the empirical adjustments available in every TPS.

\subsection{Role of the BPC}
The more significant implication of using $S_\mathrm{em}$ is operational. If we don't need $S_\mathrm{mixed}$, we don't need the BPC. That is proven by this very paper, where we measure and parameterize the halo without using a BPC. The BPC is, at most, a check on the radial integral and not, as generally claimed, essential. As a result, a substantial fraction of the halo literature to date is beside the point.   

Though we did no BPC measurements, we can confirm experimentally one of the better known facts: a 4\,cm radius would not encompass the halo at 177\,MeV. Integrating the MI fit to 4\,cm, as compared with 10\,cm, we find a BPC dose defect (maximum with $z$) of $2.8\,\%$. Lin et al. \cite{Lin2014} measure $2.6\,\%$ at 180\,MeV using radiochromic film in SolidWater (their Fig.\,9).

\subsection{The full integral of $D(r,z)$}
We have integrated the dose over $r$ from 0 to 10\,cm to obtain $S_\mathrm{mixed}(z)/\rho_\mathrm{water}$ (Fig.\,\ref{fig:S_EM}). The further integral over $z$ should equal the beam energy less the energy deposited by neutrals outside a cylinder of length (say) 23\,cm and diameter 20\,cm. We performed that integral four ways. First, recognizing that we did not really measure the halo for $z<2$\,cm, we integrated from 2 to 23\,cm and compared the answer with the nominal residual energy at 2\,cm. We also took the integral from 0 to 23\,cm and compared that with the full energy. We did each of those using both the MD and MI models. Averaging them we obtain $8.2\%$ for the energy defect $\equiv$ ((beam energy/full integral) $-$ 1), with a spread between the four methods of $\sigma=0.3\%$.

At first glance $8.2\%$ seems rather large. The energy lost to nonelastics is expected to be about 12\% \cite{Gottschalk2014} and one-third of that or 4\% is thought \cite{pedroniPencil} to go to neutrals. However, the integral is directly affected by the absolute normalization of our measurements and by the symmetry assumption, which treats $x$ as $r$. At present all we can say is that the full integral is reasonable, and that our absolute dose per proton is not far wrong. 

\subsection{Janni, nonelastic reactions, and the Bortfeld analytic formula}
As is well known, Janni \cite{janni82} tabulates the probability of a {\em nonelastic} reaction in many materials as a function of proton energy. We have now seen that  {\em elastic} scattering on H and O also contributes significantly to the halo. It is, quite properly, not included by Janni. 

That Bortfeld \cite{bortfeld} uses Janni's data may therefore account for Bortfeld's formula being slightly low around 3/4 range, as already observed by him and as parameterized by Zhang et al. \cite{Zhang2011}. Bortfeld's $D$ is, within a constant factor, our $S_\mathrm{mized}$ because he assumes a broad parallel beam. It is convenient, then, that we actually don't need $S_\mathrm{mixed}$ but rather $S_\mathrm{em}$, which can also be found from Bortfeld's formula by turning nuclear reactions off i.e. setting his parameter $\beta$ to 0.

\subsection{Our data as a Monte Carlo benchmark}
Table\;\ref{tbl:dmlg} should furnish by far the best data, to date, to test the nuclear model of a MC at an energy appropriate to proton radiotherapy. The regions dominated by EM effects and by nuclear effects are separated nearly as well as possible (a smaller beam would be better), and coherent and incoherent reactions dominate different parts of the dose distribution. Therefore defects in the MC nuclear model, if any, can not only be detected but also diagnosed to some extent.

Note that an {\em absolute} comparison is called for. Incident beam parameters, specifically $p_{1-2}$ and $p_{5-7}$, may be found in \cite{Gottschalk2014} Table\,2.

Even that part of the aura that shows is not without interest. The simulation of neutrals involves a source term and the effect of intervening materials. Here we are as close to the source as physically possible, and the geometry of intervening material (water) is trivial and well defined. Thus a simulation is a very direct test of the neutron and $\gamma$-ray source model.

\subsection{Implications for nuclear buildup}
Our investigation of the core and halo may have inadvertently solved an old problem. When a broad proton beam is sent into a water tank from air or vacuum, a dose buildup of a few percent with a buildup distance of 1--2\,cm is observed. It comes from charged secondaries, mostly protons, and was already measured in 1977 at 185\,MeV by Carlsson and Carlsson \cite{carlsson}. They noted that the magnitude of the buildup was smaller than expected.

However, their `expected' refers to conditions of transverse nuclear equilibrium, with (as we now know) a beam radius comparable to one-third the range, or $\approx7$\,cm. They did not report their beam size but assuming (as appears likely) that it was smaller, they would have observed a smaller buildup. A consequence of the core/halo picture is that nuclear buildup as measured on the axis of a finite beam by a finite detector is a function of both the beam and detector size. 

\section{Summary}
The dose distribution of a proton pencil beam stopping in water has distinctive components which we have termed {\em core}, {\em halo}, {\em aura} and {\em spray}. The first three, by definition, arise from basic physics. 

The core comes from primary protons, which have suffered only small multiple EM interactions. Except for a function describing the attrition of primaries, it is calculable from first principles given the properties of the incident beam.

The halo is caused by charged secondaries from hard single scatters be they elastic, inelastic or nonelastic. The aura is caused by neutral products (neutrons or $\gamma$\,s) of hard single scatters.

Spray is in principle avoidable. It comes from beamline components such as profile monitors, beam pipes, or degraders near the patient, acting alone or in combination. It may change with transverse beam position and may, therefore, be difficult to parameterize.

The relevant physics was studied some fifty years ago, providing the models for today's Monte Carlos. The model of the core embodies the Bethe-Bloch theory of slowing down, theories of  range straggling, and the Moli\`ere theory of MCS. 

Coherent reactions contributing to the halo include Coulomb and nuclear scattering, which obeys an optical model that explains the observed persistence of the Bragg peak to large radii. 

Incoherent scattering is described by the Bertini model augmented by models describing the decay of the residual nucleus. Knockout reactions (p,2p) and (p,pn) play a significant role. Their kinematics determine the halo radius, roughly one-third the range of the incident beam, near midrange.

We have measured the dose distribution at 177\,MeV in a straightforward manner using s single beam in a water tank. An ADCL calibrated thimble IC and a Faraday cup calibrated integral beam monitor provide an absolute measurement without renormalization. Scanning in depth, instead of radially, simplifies data taking and better reveals the form of the halo. 

Shortcomings of our measurement include positioning errors of order 1\,mm, $\pm3\%$ uncertainty in the number of incident protons, and inability to verify the radial symmetry expected under the circumstances. Despite these drawbacks our data provide by far the most comprehensive picture of the dose distribution at any single energy. Consequently, they also allow an incisive test of the MC nuclear model at a clinical energy.

We have fit our data in MD and MI fashion. Our MD form uses a purely EM stopping power in place of the widely used mixed stopping power. Our choice renders the BPC unnecessary, and improves the accuracy of the core around midrange. We do not regard either fit as the final word for clinical purposes. That will probably be a hybrid. Measurements at a number of energies will be needed to determine the energy dependence of fit parameters.

The PSI paper, along with many others, shows that their parameterization is perfectly adequate for clinical use. Even so, developers of new systems should consider our alternative as being simpler, closer to the physics, and requiring less adjustment at midrange. The improvement will be greatest for small fields. 

\section{Acknowledgements}
BG thanks Harvard University, the Physics Department, and the Laboratory for Particle Physics and Cosmology for their continuing support. We thank Drs. Grevillot, Pedroni, Sawakuchi and Zhu for communication regarding their work, and the referees for helpful comments.

\appendix

\clearpage
\section{Relation Between BPC and Broad Beam Depth Scans}\label{sec:depthScans}
We wish to show that the integrated dose $D$ to water per incident proton, measured by a BPC straddling a single pencil beam, equals the local dose to water divided by the fluence in air on the axis of a sufficiently broad  uniform beam. 

\begin{figure}[p]
\centering\includegraphics[width=4.095in,height=3.5in]{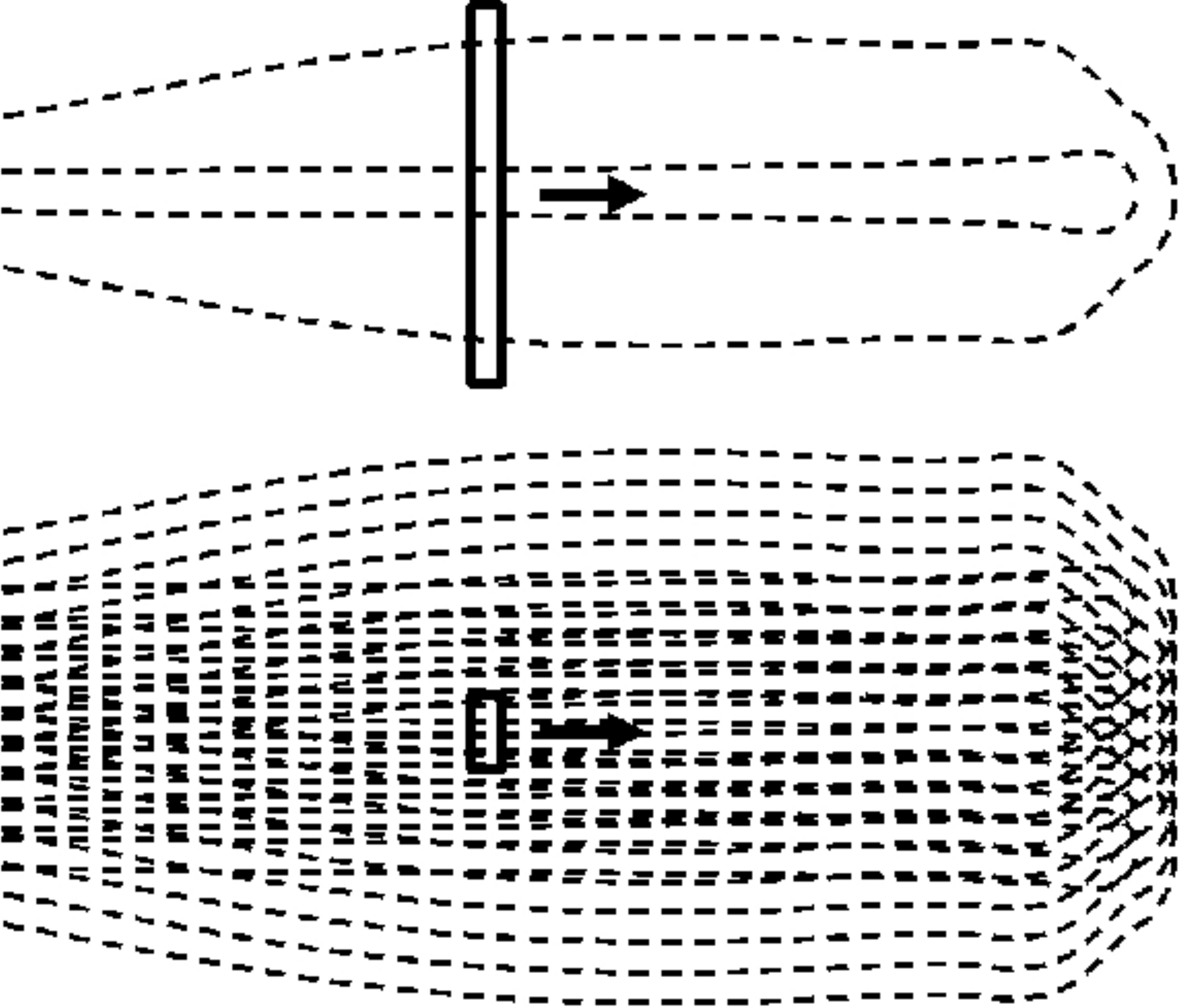}% 1.17
\caption{Top: depth scan of a single pencil beam with a Bragg peak chamber (BPC). Bottom: depth scan of a broad beam with a small IC.\label{fig:depthScans}}
\end{figure}

The average energy $dE_1$ per proton deposited in a disk of thickness $dz_1$ and radius $R_\mathrm{halo}$ by many protons entering a large water tank near the axis (cf. Fig.\,\ref{fig:depthScans} top) is
\begin{equation}
dE_1(z)\;=\;\left(\int_0^{2\pi}\int_0^{R_\mathrm{halo}}D_1(r,z)\;r\;dr\;d\phi\right) \rho\;dz_1
\end{equation}
Now let the tank be exposed, instead, to a broad uniform parallel proton beam of fluence in air $\Phi$. The energy $dE_n(z)$ deposited in a small disk of thickness $dz_1$ and radius $R_\mathrm{dosim}$ by $n$ protons at the tank entrance directed at the disk (cf. Fig.\,\ref{fig:depthScans} bottom) is
\begin{equation}
dE_n(z)\;=\;\left(\int_0^{2\pi}\int_0^{R_\mathrm{dosim}}D_n(r,z)\;r\;dr\;d\phi\right) \rho\;dz_1
\;=\;\pi\;R_\mathrm{dosim}^2\;D_n(0,z)\;\rho\;dz_1
\end{equation}
In the second step we have assumed transverse equilibrium (EM and nuclear) near the axis, so that $D_n$ is independent of $r$ there. Because the mix of stopping powers is the same in both discs, and they have the same thickness, it follows that
\begin{equation}\label{eqn:dEn}
dE_1(z)\;=\;dE_n(z)/n
\end{equation}
Transverse equilibrium also plays a role in Eq.\,\ref{eqn:dEn}. Protons initially directed at the small dosimeter do not deposit all their energy in it. However, the shortfall is exactly compensated by energy from protons {\em not} directed at the dosimeter. We thus find
\begin{equation}
\int_0^{2\pi}\int_0^{R_\mathrm{halo}}D_1(r,z)\;r\;dr\;d\phi\;=\;D_n(0,z)/\bigl(n/(\pi\,R_\mathrm{dosim}^2)\bigr)
  \;=\;D_n(0,z)\,/\,\Phi
\end{equation}
Dose per fluence is a mass stopping power, which we have called $S_\mathrm{mixed}/\rho_\mathrm{water}$ .
\clearpage

%%%%%%%%%%%%%%%%%%%%%%%%%%%%%%%%%%%%%%%%%%%%%%%%%%%%%%%%%%%%%%%%%%%%%%%%%%%%%%%%%%%%%%%%%%%%%%%%%%%%%%%%%%%%%%%%%%%%%%%%%%%%%%%%%%%%%%%%%%
\begin{table}[h]
\setlength{\tabcolsep}{5pt}
\begin{center}
\begin{tabular}{rrrrrrrrrrr}
\multicolumn{1}{c}{z/r}&           
\multicolumn{1}{c}{0.0}&           
\multicolumn{1}{c}{1.0}&           
\multicolumn{1}{c}{2.0}&           
\multicolumn{1}{c}{3.0}&           
\multicolumn{1}{c}{4.0}&           
\multicolumn{1}{c}{5.0}&           
\multicolumn{1}{c}{6.0}&           
\multicolumn{1}{c}{7.0}&           
\multicolumn{1}{c}{8.0}&           
\multicolumn{1}{c}{10.0}\\           
 1.533&       M& -0.4274&       M& -2.7179& -3.3276& -3.7252&       M&       M& -4.2708& -4.4897\\
 2.000&  0.4398& -0.3994& -1.9724& -2.5946& -3.2508& -3.6650& -3.8861& -4.1493& -4.2596&       M\\
 3.000&  0.4406& -0.3460& -1.8342& -2.3751&       M&       M& -3.7725&       M&       M& -4.4656\\
 4.000&  0.4438& -0.3267& -1.7684& -2.2437& -2.8720& -3.3366& -3.6435& -4.0060& -4.1789&       M\\
 5.000&  0.4444& -0.3116& -1.7244& -2.1716&       M&       M& -3.5098&       M&       M& -4.4177\\
 6.000&  0.4465& -0.2945& -1.6886& -2.1196& -2.6220& -3.0377& -3.3674& -3.7823& -4.0284&       M\\
 7.000&  0.4471& -0.2796& -1.6509&       M&       M&       M& -3.2603&       M&       M& -4.3311\\
 8.000&  0.4477& -0.2603& -1.6184& -2.0575& -2.5106& -2.8524& -3.1636& -3.5861& -3.8656&       M\\
 9.000&  0.4476& -0.2427& -1.5831&       M&       M& -2.8031& -3.0898&       M&       M& -4.2184\\
10.000&  0.4490& -0.2215& -1.5479& -2.0126& -2.4641& -2.7707& -3.0398& -3.4558& -3.7332&       M\\
11.000&  0.4506& -0.2025& -1.5114&       M&       M& -2.7500& -3.0099&       M&       M& -4.1351\\
12.000&  0.4504& -0.1787& -1.4714& -1.9694& -2.4401& -2.7402& -2.9951& -3.3940& -3.6628&       M\\
13.000&  0.4519& -0.1558& -1.4286&       M&       M& -2.7382& -2.9944&       M&       M& -4.0866\\
14.000&  0.4534& -0.1282& -1.3798& -1.9216& -2.4234& -2.7417& -3.0054& -3.3838& -3.6381&       M\\
15.000&  0.4566& -0.1000& -1.3279&       M&       M& -2.7546& -3.0280&       M&       M& -4.0727\\
16.000&  0.4618& -0.0690& -1.2657& -1.8632& -2.4153& -2.7726& -3.0690& -3.4454& -3.6837& -4.1002\\
17.000&  0.4801& -0.0300& -1.1926& -1.8213& -2.4084& -2.8035& -3.1125& -3.4600&       M& -4.1740\\
18.000&  0.4915&  0.0169& -1.1083& -1.7703& -2.3906& -2.8041& -3.1075& -3.4741& -3.7672& -4.2738\\
18.500&  0.5046&  0.0459& -1.0593& -1.7321& -2.3659& -2.7881& -3.1002& -3.4997&       M&       M\\
19.000&  0.5232&  0.0802& -1.0022& -1.6839& -2.3317& -2.7638& -3.0866& -3.5121& -3.8568& -4.3552\\
19.200&       M&       M&       M& -1.6602&       M&       M&       M& -3.5142&       M&       M\\
19.400&       M&       M&       M& -1.6348&       M&       M&       M& -3.5226&       M&       M\\
19.500&  0.5510&  0.1253& -0.9328&       M& -2.2810& -2.7112& -3.0350&       M& -3.9701&       M\\
19.600&       M&       M&       M& -1.6076&       M&       M&       M& -3.5468&       M&       M\\
19.800&       M&       M&       M& -1.5757&       M&       M&       M& -3.5921&       M&       M\\
20.000&  0.5993&  0.1898& -0.8433& -1.5368& -2.1845& -2.6303& -3.0308& -3.6658& -4.1091& -4.3940\\
20.200&  0.6261&  0.2266& -0.7987& -1.4900& -2.1520& -2.6292& -3.0882& -3.7601& -4.1539&       M\\
20.400&  0.6662&  0.2741& -0.7438& -1.4443& -2.1500& -2.6695& -3.1958& -3.8699& -4.1808&       M\\
20.600&  0.7169&  0.3306& -0.6859& -1.4146& -2.1974& -2.7669& -3.3555& -3.9737&       M&       M\\
20.800&  0.7625&  0.3765& -0.6353& -1.4176& -2.3110& -2.9298& -3.5583& -4.0469&       M&       M\\
21.000&  0.7747&  0.3835& -0.6096& -1.4712& -2.5038& -3.1666& -3.7683& -4.0937& -4.2238& -4.3924\\
21.200&  0.7208&  0.3199& -0.6364& -1.5919& -2.7751& -3.4568& -3.9238&       M&       M&       M\\
21.400&  0.5714&  0.1605& -0.7491& -1.8058& -3.1200& -3.7335& -3.9970&       M&       M&       M\\
21.500&       M&       M&       M&       M&       M&       M&       M& -4.1368&       M&       M\\
21.600&  0.3094& -0.1285& -0.9806& -2.1316& -3.4818& -3.8935& -4.0253&       M&       M&       M\\
21.800& -0.0945& -0.5719& -1.3676& -2.6019& -3.7381& -3.9465& -4.0363&       M&       M&       M\\
22.000& -0.6658& -1.2025& -1.9513& -3.1835& -3.8593& -3.9629& -4.0462&       M& -4.2504& -4.4152\\
22.200& -1.4468& -2.0547& -2.7337& -3.6274&       M&       M&       M&       M&       M&       M\\
22.400& -2.4194&       M& -3.4703& -3.7514&       M&       M&       M&       M&       M&       M\\
22.500&       M&       M&       M&       M& -3.8977& -3.9856&       M&       M&       M&       M\\
22.600& -3.3715& -3.5933& -3.6934& -3.7737&       M&       M&       M&       M&       M&       M\\
22.800&       M& -3.6613& -3.7298&       M&       M&       M&       M&       M&       M&       M\\
23.000&       M&       M& -3.7546& -3.8084& -3.9206& -4.0035& -4.0735&       M& -4.2678& -4.4219\\
23.500&       M&       M& -3.8008& -3.8496&       M&       M&       M&       M&       M&       M\\
24.000&       M&       M& -3.8456& -3.8863& -3.9752& -4.0544& -4.1141&       M&       M& -4.4374\\
25.000&       M&       M&       M&       M&       M&       M&       M&       M&       M& -4.4517\\
\multicolumn{1}{c}{z/r}&           
\multicolumn{1}{c}{0.0}&           
\multicolumn{1}{c}{1.0}&           
\multicolumn{1}{c}{2.0}&           
\multicolumn{1}{c}{3.0}&           
\multicolumn{1}{c}{4.0}&           
\multicolumn{1}{c}{5.0}&           
\multicolumn{1}{c}{6.0}&           
\multicolumn{1}{c}{7.0}&           
\multicolumn{1}{c}{8.0}&           
\multicolumn{1}{c}{10.0}\\           
\end{tabular}
\end{center}
\caption{Measured log$_{10}$(dose/(MeV/g/p)) at various depths $z$ (cm) and distances $r$ (cm) from the beam centerline.\label{tbl:dmlg}}
\end{table}
\clearpage

%%%%%%%%%%%%%%%%%%%%%%%%%%%%%%%%%%%%%%%%%%%%%%%%%%%%%%%%%%%%%%%%%%%%%%%%%%%%%%%%%%%%%%%%%%%%%%%%%%%%%%%%%%%%%%%%%%%%%%%%%%%%%%%%%%%%%%%%%%
\listoffigures
\clearpage

\clearpage

%\bibliographystyle{unsrt}
%\bibliography{/pctexv4/work/pbs/master}

\end{document}